\documentclass[runningheads]{llncs}

\usepackage{eccv}

\usepackage{eccvabbrv}

\usepackage{graphicx}
\usepackage{booktabs}
\usepackage{enumitem}
\usepackage{multirow}
\usepackage{wrapfig}
\usepackage{overpic}

\usepackage[accsupp]{axessibility}  % Improves PDF readability for those with disabilities.

\usepackage[pagebackref]{hyperref}
\usepackage[table]{xcolor}
\usepackage{orcidlink}
\usepackage{svg}
\usepackage{amsmath}
\usepackage{makecell}
\usepackage{wrapfig}
\usepackage{tikz}

\definecolor{lightboldcolor}{gray}{0.6}

\begin{document}

\title{Geometry Gaussians: Decoupling Appearance and Geometry in Gaussian Splatting} 
%\title{Beyond Appearance: A Dual-Opacity Representation for Geometry-Aware Gaussian Splatting}

\titlerunning{Abbreviated paper title}

\author{Hongyu Zhou\inst{1,2}\orcidlink{0000-0002-0099-643X} \and Zorah Lähner\inst{1,2}\orcidlink{0000-0003-0599-094X}}

\authorrunning{H.Zhou, Z.Lähner}

\institute{University of Bonn, Germany %\\
%\email{\{hzhou,laehner\}@uni-bonn.de}
\and
Lamarr Institut
}

\maketitle

\begin{figure}
    \centering
    \begin{overpic}[trim=0.0cm 5.9cm 2.7cm 0.0cm, clip,width=0.95\linewidth]{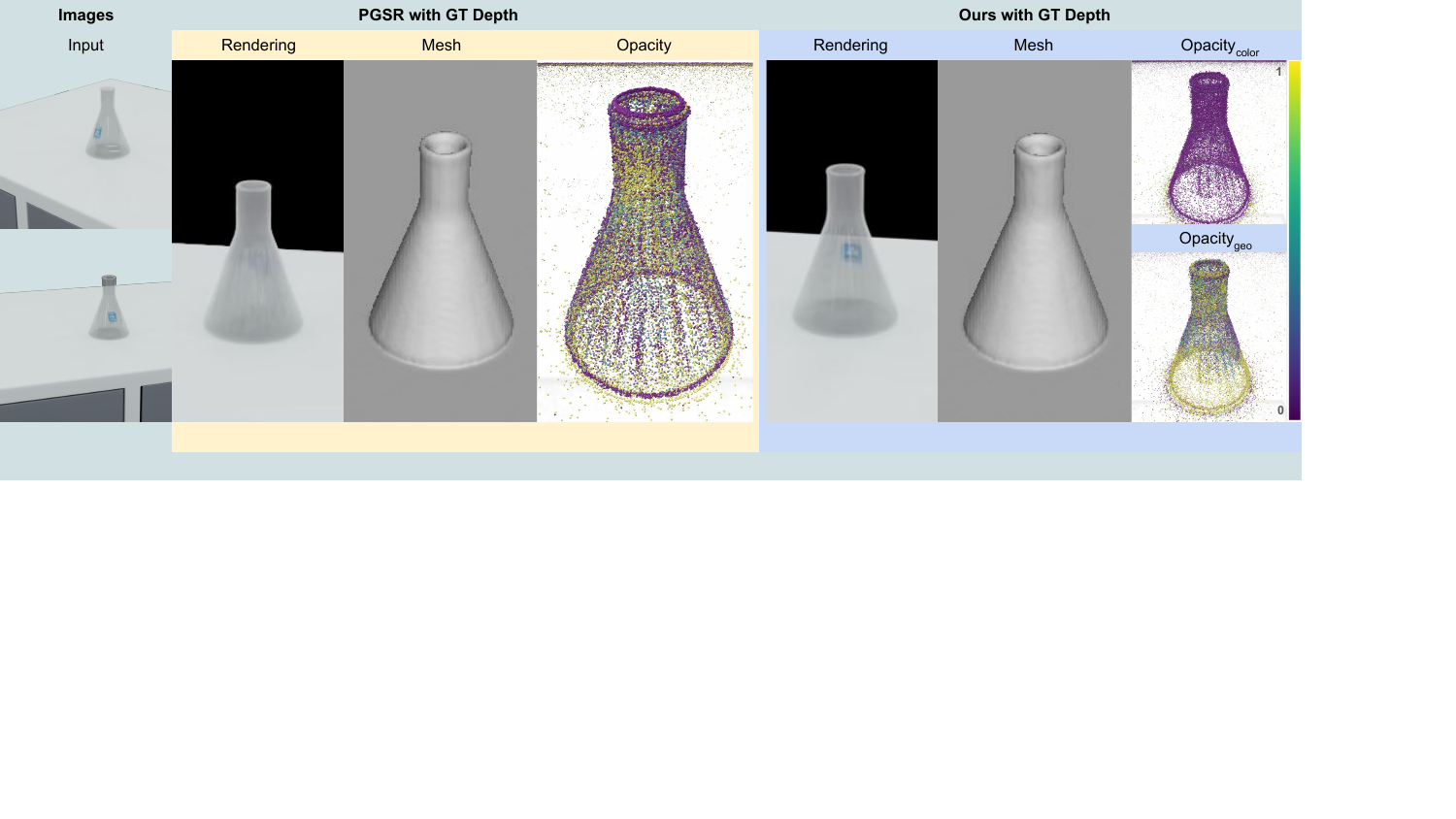}\put(20,17){\color{red}\circle{3}}\end{overpic}
    \caption{Comparison of 3DGS trained with full knowledge of the texture and geometry (in form of depth maps) without and with our method. A single additional parameter opacity$_{geo}$ for each splat allows to capture the full scene information by separating color from geometry information which is especially important for transparent objects.  }
    \label{fig:teaser}
\end{figure}

\begin{abstract}
    After the success of 3D Gaussian Splatting (3DGS) for novel view synthesis, many works have explored how to also use it for geometric surface representation. 
    However, extracting accurate geometric information directly from 3DGS remains challenging and can often reduce the appearance rendering quality.
    In this work, we show that 3DGS in its default form is inheritedly unsuited to represent texture and geometry at the same time, by training with complete ground-truth texture and geometry information.
    We also propose a simple solution by applying a single additional geometry opacity parameter to each splat, together with an optional transparency-curated optimization pipeline. 
    Our experiments, both with ground-truth and vision foundation model geometric input, show that this change leads to improved rendering and geometry performance on a wide variety of dataset, and especially complex scenes with transparent objects benefit significantly from our method. 
    %\keywords{3D Gaussian Splatting \and Geometric Representation}
\end{abstract}    
\section{Introduction}\label{sec:intro}

\begingroup
3D Gaussian Splatting (3DGS)~\cite{kerbl20233d} has introduced, with great success, a new representation for novel view synthesis that enables high-quality, real-time rendering and efficient training. 
Due to its wide adaption, there is a growing interest in using 3DGS for other applications, including those that need a faithful 3D geometry, for example robotics, augmented reality, or physical simulations~\cite{zhu20243d, yu2025artgs, zhai2025splatloc,cao2026physgaussian}.

However, a fundamental tension exists in the core of Gaussian Splatting: the splats that best explain the appearance of a scene are not necessarily those that best describe its geometry. 
This mismatch is intuitively clear for scenes containing transparent or translucent objects, for which the color originates at least partially from behind the object but the surface is not diffuse, see \cref{fig:transparency}.
\setlength\intextsep{0pt}
\begin{wrapfigure}[20]{r}{0.4\textwidth}
\begin{overpic}[width=\linewidth]{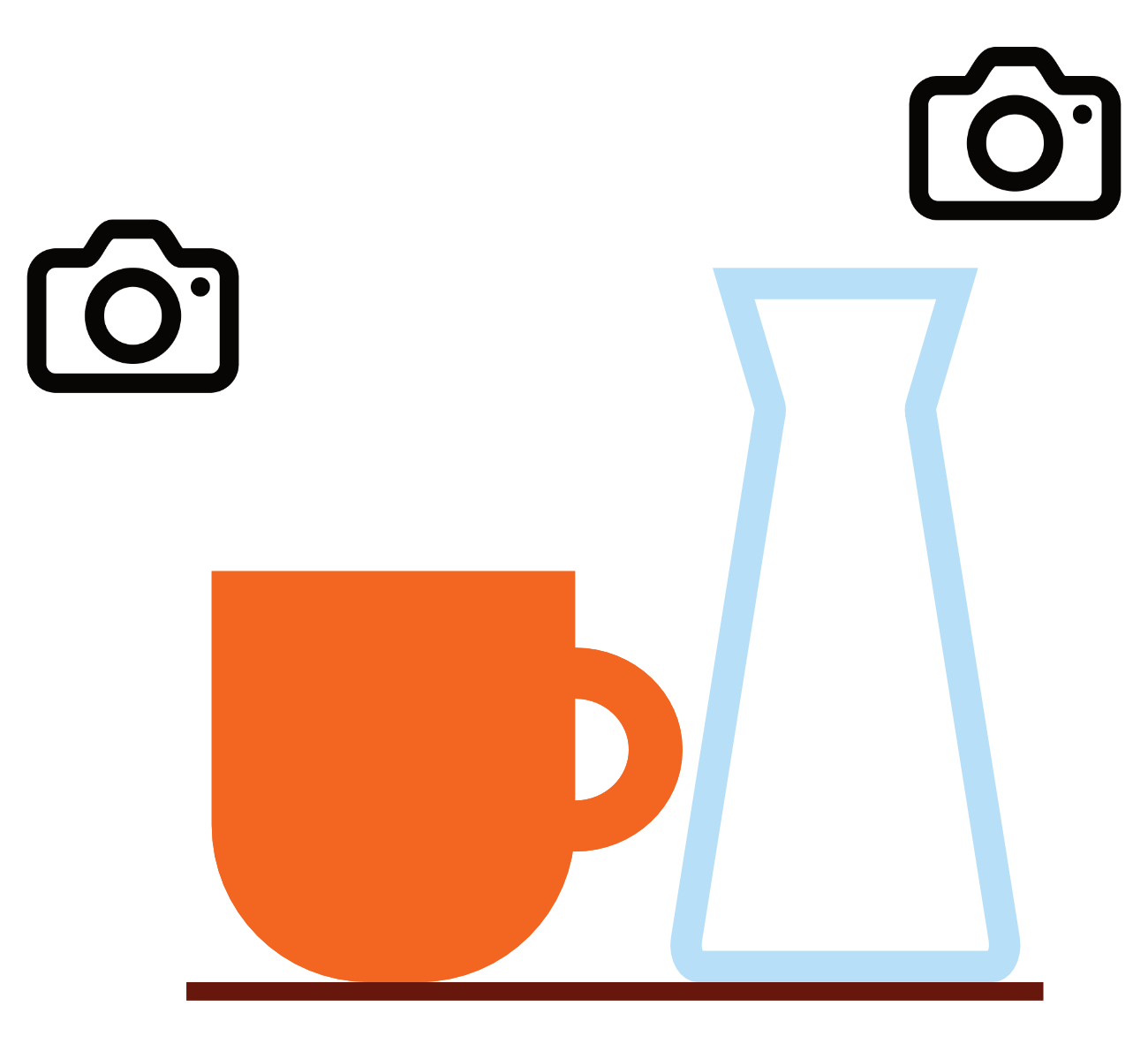} 
    \put(20,17){$A$}
    \put(18.5,21){\circle*{3}}
    \put(5.5,34){\thicklines\vector(1,-1){12}}
    \put(42,33){$C$}
    \put(45,41){\circle*{3}}
    \put(25,61){\thicklines\vector(1,-1){19.5}}
    \put(75,71){\thicklines\vector(-1,-1){29}}
    \put(83,42){$B$}
    \put(82,47){\circle*{3}}
    \put(92,57){\thicklines\vector(-1,-1){32.5}}
\end{overpic}
\caption{(A) In most objects color and geometry of a view locate at the same point. (B) For transparent objects, color is determined by objects behind the geometry. (C) If the object behind is seen from a different view point, the color rendering through transparency can be accurately represented. }
\label{fig:transparency}
\end{wrapfigure}
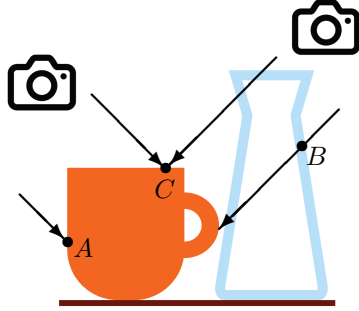 
Existing geometry extraction methods rely on opacity thresholds~\cite{yu2024gaussian}, or simultaneous reconstruction with another geometry representation~\cite{guedon2025milo,zhu2025gaussian} but still depends heavily on the assumption that visual contribution and geometric occupancy are tightly related. 
We argue, and experimentally validate in \cref{sub:failure}, that this is a representational shortcoming of how Gaussian Splatting is commonly implemented but can be overcome with minimal adjustment. 
In standard 3DGS, each splat has a single opacity parameter that is used for both its contribution to rendering as well as for the scene geometry. 
We propose to decouple these roles by equipping each splat with an additional \emph{geometry opacity}: a learned scalar that is only responsible for geometric quantities, such as depth and normals. Similar architecture has been explored in \cite{li2025car} but they are limited to the reconstruction task of cars.
We further claim that this simple change has a significant positive effect on both the rendering and geometry reconstruction while leaving large parts of the 3DGS pipeline untouched. 
The additional geometry opacity parameter is not included in parts that only concern color rendering, but replaces the existing opacity in geometric regularization. 
This only adds a single scalar per primitive and is immediately compatible with any existing geometry-aware 3DGS pipeline.

We validate our geometry opacity on multiple datasets with different materials and see consistent improvements for complex settings.
Our method is particularly effective on transparent objects, where previous methods often either miss large parts of the geometry, or introduce artifacts in rendering~\cite{li2025tsgs}, while ours recovers a coherent object and improve the rendering at the same time. 
Despite its simplicity, we believe this explicit decoupling of rendering opacity and geometry opacity represents an important step towards Gaussian Splatting representations that are faithful in both appearance and structure, and thus applicable in a wide range of applications.
\endgroup
    
\paragraph*{Contributions. } The key contributions of our work are the following:
\begin{itemize}[label=\textbullet]
    \item We show that jointly encoding rendering and geometry information in the geometry-aware 3DGS frameworks leads to conflicting information and, thus, suboptimal results; especially for rendering which is less regularized. 
    \item With \emph{geometry opacity} parameter for each splat, color rendering and geometry can be cleanly separated without introducing considerable computational or memory overhead. We show that it can improve the reconstruction without reducing rendering quality when applying geometric regularization.
    \item We introduce a practical pipeline to include geometry opacity in combination with vision foundation models. It includes several regularizers aimed at transparent objects which benefit significantly from the geometric disentanglement, including a transparency-aware multi-view stereo loss.
    \item Our experiments validate influence of geometry opacity, especially for challenging cases with transparent objects. We achieve state-of-the-art results on the datasets NeRF Synthetic, DTU, TransLab and Mip-NeRF. 
\end{itemize}
% The code and complete experimental setup will be published after acceptance.

\section{Related Work} \label{sec:relatedwork}

\subsection{General 3DGS}
3D Gaussian Splatting~\cite{kerbl20233d} is an efficient novel view synthesis method representing the scene as a sparse collection of 3D Gaussian distributions with color and opacity information.
Even though it is fairly recent, it has been widely adapted to different applications~\cite{SplattingAvatar:CVPR2024,yan2024street,yan2023gsslam}, made even more efficient~\cite{chen2024mvsplat}, and combined with other representations, like signed distance functions, to produce more accurate geometry~\cite{guedon2023sugar,yu2024gsdf,lyu20243dgsr,Huang2DGS2024}. Another line of work, for example PGSR~\cite{chen2024pgsr}, aims for a better estimation of the depth and normal maps through single- and multi-view regularization during optimization.

However, while non-Lambertian effects are well captured in the view-dependent rendering equation, geometric reconstruction from 3DGS often struggles with specular effects, leading to artifacts~\cite{yao2025reflective}. 
This is because volume rendering does not capture surface properties well in these cases.
Directly regularizing (approximated) surface properties like normals improves the reconstruction by providing additional guidance in geometric domains~\cite{yu2024gaussian}.
Another solution is using separate geometric priors predicted directly from images trained on large image datasets.
Due to the learned prior knowledge, this produces reliable information about the surface even under very inconsistent visual cues or illumination settings~\cite{wang2025gsi3}.
Similar ideas have been explored for transparent surfaces, for example by predicting a rendering of the same object with matte material properties~\cite{li2025tsgs}.

\subsection{Geometric Representation in 3DGS}
SuGaR~\cite{guedon2023sugar} pioneered geometric regularization in 3DGS by encouraging splats to lie on the surface of a jointly extracted mesh.
Afterwards, more and more work starts to add geometric regularization to train Gaussian Splatting. 
However, a trade-off between rendering and reconstruction can be seen in many works, \eg Gaussian Opacity Field~\cite{yu2024gaussian}, 2D Gaussians~\cite{Huang2DGS2024}, GSDF~\cite{yu2024gsdf}, and PGSR~\cite{chen2024pgsr}. 
CarGS~\cite{shen2025evolving} decouples geometry and rendering by learning an additional covariance matrix for geometry, improving the performance on both aspects, but they fail to capture transparent objects. 
Zhen et al.~\cite{tan2025uncertainty} learns an uncertainty coefficient for each Gaussian, but they focus on sparse-view reconstruction and this does not decouple the geometry. 
For reflection, Ref-Unlock~\cite{song2025reflections} learns transmittance and reflection components separately, and physical-based rendering methods~\cite{jiang2024gaussianshader, liang2024gs} learn BRDF values on Gaussians to model them better, but these methods do not focus on reconstruction and the positive effects do not extend to ordinary scenes. Car-gs~\cite{li2025car} introduce hybrid opacity to render geometry, but they solely focus on 3D reconstruction of cars.
In this paper, unlike car-gs~\cite{li2025car}, we focus on the trade-off between the performance on reconstruction and rendering and we show that the geometric opacity is beneficial for both sides with our proposed optimization.
%We render geometric features via the geometric opacity and we are able to increase both reconstruction and rendering quality. 
% This allows our method to model the transmittance and is especially useful for transparent objects.
% \begin{itemize}
%     \item 2D Gaussians \cite{Huang2DGS2024}
%     \item SuGAR \cite{guedon2023sugar}
%     \item GOF \cite{yu2024gaussian}
%     \item NeuSG \cite{chen2025neusgneuralimplicitsurface}
%     \item that different covariances work \cite{shen2025evolving}
%     \item GSDF
% \end{itemize}

\subsection{Vision Foundation Models}
Pretrained models for predicting geometric priors, like depth and normals, from images have become quite common. 
StableNormal~\cite{ye2024stablenormal} is able to produce monocular normal estimation from a single image. Depth Anything~\cite{yang2024depthanything} and MuGe~\cite{wang2025moge} shows that nowadays accurate depth estimation is possible for almost all everyday scenes. 
In theory monocular depth estimation is an ill-posed problem mathematically but due to large training sets neural networks can learn scale priors that allow meaningful predictions anyway~\cite{Ranftl2022}. 
Nowadays, more effort has been put into \emph{multi-view} geometry estimation. 
MASt3R~\cite{leroy2024grounding} and DUSt3R~\cite{wang2024dust3r} provide multi-view depth estimation, but they rely on pair-wise estimation that requires post-processing for further usage. 
More recently, VGGT~\cite{wang2025vggt} and Depth Anything 3~\cite{lin2025depth} enabled multi-view depth estimation and point estimation through one single forward pass. 
While effective for open-domain images, they struggle with semi-transparent objects. 
DKT~\cite{xu2025diffusion} fine-tuned a large video diffusion model on data with transparency to build a perception model for these cases, however the accuracy for normal scenes decreases. 
Fine-tuning a large foundation model without damaging its original performance can be difficult. 
Thus, in this paper we propose using multiple models and assigning them the task they excel at.

\subsection{Geometric Supervision for 3DGS}
Many methods have proposed to integrate geometric priors into NeRF or Gaussian Splatting. 
Mvpgs~\cite{xu2024mvpgs} explores multi-view priors for Gaussian Splatting, but they are limited to rendering and do not study into 3D reconstruction. 
On the other hand, MVG-splatting~\cite{li2025mvg} uses multi-view priors for 3D reconstruction. % it said "does not have good performance" here, thats a bit too harsh ;) 
MonoSDF~\cite{yu2022monosdf} uses monocular depth and normal priors to supervise the SDF training for 3D reconstruction but monocular predictions are often inconsistent from different views. 
GausSurf~\cite{wang2024gaussurf}, TSGS~\cite{li2025tsgs}, Reflections Unlock~\cite{song2025reflections} and 2DGS-Room~\cite{zhang20242dgs} also leverage monocular estimation to train Gaussian Splatting for more complex scenarios, and MILo~\cite{guedon2025milo} uses geometric priors to supervise the joint training of Gaussian Splatting and a mesh. 
DepthSplat~\cite{xu2025depthsplat} focuses on the latent space and learns depth estimation alternating with Gaussians, but its feed-forward method is orthogonal to our work.

In our work, we follow the idea of using foundation models to predict pseudo geometric ground truth, but we acknowledge explicitly that the predictions can be erroneous and develop a learning-in-cycle mechanism for joint optimization of both geometric estimation and Gaussian Splatting to overcome this.
\section{Geometric Representation} \label{sec:geometricrepresentation}

In \cref{sub:failure} we will show that the default 3DGS parametrization cannot fully capture texture and geometry information in a scene and propose a small but effective change to the parametrization in \cref{sub:geoopacity} to overcome this lack in representation. 

\subsection{Limited Representation Power in 3DGS} \label{sub:failure}

Extracting surface information directly from 3DGS is challenging because the splats have been shown to not accurately localize on the surface for higher frequency texture~\cite{shen2025evolving}.
This can be improved by adding geometric information, in the form of depth and normal maps generated by VGGT~\cite{wang2025vggt}, to the supervision signal.
However, these solutions extract geometry from the same set of splats.
But there are cases with transparency in which surface and color information do not come from the same points, which cannot be modeled by standard 3DGS. 

We conduct a series of experiments in which we provide dense ground-truth color and geometry information to two GS methods focused on geometric reconstruction.
The results can be seen in \cref{exp:quantitative:representation power} and \cref{exp:qualitative:representation power}.
Neither method can accurately capture geometry \emph{and} rendering in complicated cases -- indicating that this is a lack of representation power in 3DGS since they could theoretically overfit on the training data. 
The reference methods perform better in geometric reconstruction than rendering probably due to more regularization focused on this. 
We strengthen this interpretation in the next section by adding minimal additional degrees of freedom, which immediately solve the problem.
% 

%\begin{itemize}
%    \item surface information is hard to be directly extracted from 3DGS
%    \item besides methods that keep a separate geometry representation, like a mesh, the geometry can be seen by rendering the depth and normals from different views, similar to image views
%    \item these can also be used as supervising signals during training (e.g. by taking depth predictions from vision foundation models)
%    \item when providing ground-truth color and geometry information for (complicated?) scenes, SOTA 3DGS methods still fail to encode both correctly
%    \item short discussion about possible reasons
%\end{itemize}

\subsection{Geometry Opacity}\label{sub:geoopacity}

We believe this shortcoming is grounded in conflicting demands for color and surface information. 
Color renderings have been shown to peak when Gaussians can be placed with not directly on but in front of the surface to model advanced texture effects.
Geometric information needs to be localized exactly on the surface to provide accurate depth and normals, which are naturally sensitive to position noise. 
It is not possible to separate these two in the default 3DGS splatting pipeline because, even though opacity to turn a splat influence lower exists, it influences color and geometry in the same way. 
To solve this, we propose a minimal change to the formulation by adding an single parameter to each splat: the \emph{geometry opacity}.
This solution disentangles the conflict without introducing much overhead. 
To show the effectiveness, we repeat the experiments from \cref{sub:failure} using complete ground-truth information while training but with included geometry opacity parameter, see \cref{exp:quantitative:representation power} and \cref{exp:qualitative:representation power}.
The results clearly show improved errors for both rendering and geometry. 
In the next section, we will propose a new framework to incorporate the geometry opacity when no ground-truth geometric information is available for the scene. 

%\begin{itemize}
%    \item We believe this shortcoming is grounded in conflicting demands for color and surface information. 
%    \item Color renderings have been shown to peak when Gaussians can be placed with not directly on but in front of the surface to model advanced texture effects.
%    \item Geometric information needs to be localized exactly on the surface to provide accurate depth and normals (which are quite sensitive to position noise). 
%    \item In existing optimizations these Gaussians on the surface exist but additional ones only needed for rendering "in space around" also exist. 
%    \item We propose a new parameter geometric opacity which encodes this information. 
%    \item This solution disentangles the conflict without introducing much overhead. 
%     \item When fitting 3DGS with ground-truth information, scenes can be represented perfectly. 
%   \item The representation is in theory superior, in the next section we show how to use it in practice. 
%\end{itemize}
\begin{figure}[tbh]
\vspace{-45pt}
    \centering
    \includegraphics[width=1.0\textwidth]{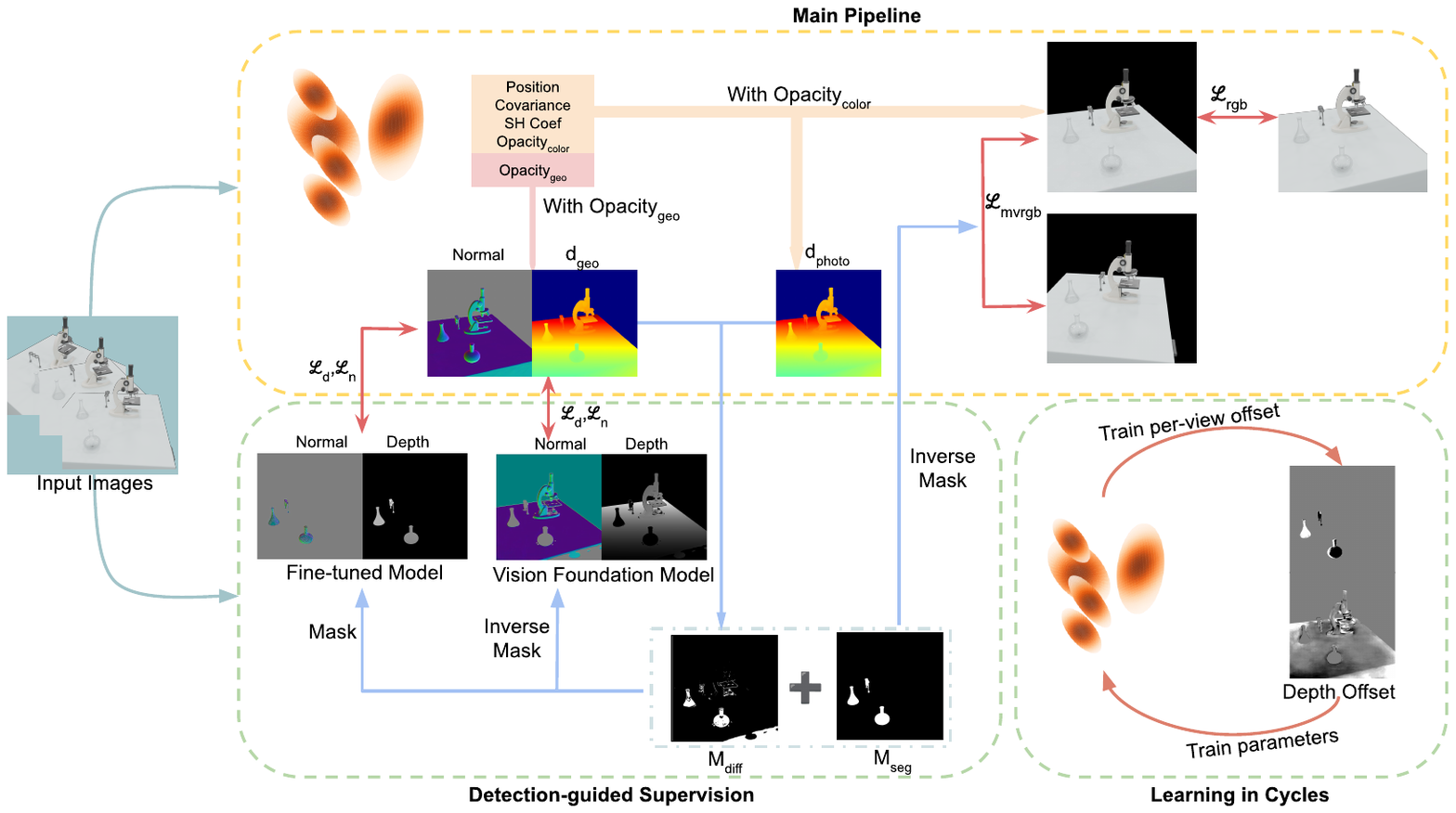}
    \vspace{-45pt}
    \caption{\textbf{Overview}. We add a new parameter opacity$_{geo}$ to each splat which is used to render depth and normal maps while the default opacity is now opacity$_{color}$ and only responsible for rendering the RGB images. The geometric part is optimized by supervision from vision foundation models, together geometric regularization from PGSR~\cite{chen2024pgsr}. In addition, we segment transparent parts that serve for masking in the geometric regularization for detection-guided supervision~(\cref{sub:detectionguidedsupervision}). Learning in cycles~(\cref{sub:learningincycles}) optimizes an depth offset for the predicted depth that overcomes small misalignment. }
    \label{fig:overview}
    \vspace{-25pt}
\end{figure}

\section{Optimization} \label{sec:method}

In real-world applications ground-truth information is not available, but can be acquired from large vision foundation models. 
However, the results are often noisy, slightly misaligned, non-metric or struggle with complex materials.
In this section, we first introduce the existing geometry-aware 3DGS optimization (\cref{sub:geometryaware}) and then three strategies to improve the geometric supervision under noise and for transparent objects in particular (\cref{sub:detectionguidedsupervision}, \cref{sub:learningincycles}, \cref{sub:transparencyawaremvs}). 
\cref{fig:overview} shows an overview of the whole pipeline. 

\subsection{Geometry-Aware Gaussian Splatting} \label{sub:geometryaware}
Depth estimation and normal estimation have been widely used in enhancing the quality of 3D reconstruction for volumetric representations as NeRF and Gaussian Splatting. In general, we follow conventions of PGSR~\cite{chen2024pgsr} and geometric supervision is applied by geometric regularization in the form of 
\begin{align}
\begin{aligned}
\mathcal{L}_{n} = \lambda_{normal}\|N_r - N_{gt}\| + \lambda_{normal}\|N_d - N_{gt}\|, \label{eq:ln}
\end{aligned}
\end{align}
and 
\begin{align}
\begin{aligned}
\mathcal{L}_{d} = \lambda_{depth}\|d_r - d_{gt}\| \label{eq:ln}
\end{aligned}
\end{align}
where $N_r$ denotes the normals rendered from Gaussians as in PGSR, $N_d$ the normals derived from the depth map, $d_r$ denotes the unbiased depth estimation~(plane depth in PGSR), and $N_{gt}, d_{gt}$ the predicted normal and depth map from foundation models, respectively. Detailed definitions can be found in \cite{chen2024pgsr}. 

PGSR uses a single-view and a multi-view loss term to regularize the geometry without using vision foundation models. They can be written as 
% \begin{align}
% \begin{aligned}
% \mathcal{L}_{svgeom} = \frac{1}{W}\sum_{p\in W}\|N_d - N_r\|, \label{eq:lsvgeom}
% \end{aligned}
% \end{align}
% and 
\begin{equation}
\mathcal{L}_{svgeom} = \frac{1}{W}\sum_{p\in I}\|N_d - N_r\|,  \hspace{4pt}\\
\mathcal{L}_{mvgeom} = \frac{1}{V}\sum_{p_r\in I}\|p_r - H_{nr}H_{rn}p_r\| \label{eq:lgeom}
\end{equation}
where $I$ is the pixel space, $p_r$ is the pixel in view $r$ and $H_{nr}$ is the homography matrix from view $n$ to view $r$. In this paper, we will adopt the above form of geometric regularization with our own changes explained in the next sections.
% \begin{equation} 
%     \mathcal{L}_{mvgeom} = \frac{1}{V}\sum_{p_r\in W}\|p_r - H_{nr}H_{rn}p_r\| \label{eq:lmvgeom}
% \end{equation} 

% \ZL{we might need to include a very short intro do geometry aware 3DGS here (or in related work), how depth is usually used for supervision}

%\begin{itemize}
%    \item while representation power is higher, ground-truth information is not usually available
%    \item depth from vision foundation models is visually good, but in reality is often noisy/non-metric/sucks in other ways
%    \item robustness to this misinformation needs to be built in the optimization
%\end{itemize}

\subsection{Detection-guided Supervision} \label{sub:detectionguidedsupervision}
Vision foundation models trained for open-domain images have improved significantly in recent years. 
However, no matter whether they are monocular as Stable Normal~\cite{ye2024stablenormal} and MoGe~\cite{wang2025moge}, or multi-view as VGGT~\cite{wang2025vggt} and Depth Anything3~\cite{lin2025depth}, their performance on transparent objects is quite unstable and inaccurate, especially in case of high transmittance material. 
While it is possible to fine-tune the model on transparent materials, as implemented in DKT~\cite{xu2025diffusion}, the performance on general objects degrades at the same time. See examples in the supplementary. 

To get the best of both worlds, we propose treating transparent objects and non-transparent objects with a model that is specialized for the case. 
Before extracting the geometric priors, we use an open-vocabulary segmentation model, in our case SAM3~\cite{carion2025sam}, to detect transparent objects in the inputs, denoted as $M_{seg}$. 
Then, instead of relying on one single visual foundation model, parts with non-transparent objects are processed with an open-domain-trained model and parts with transparency use a model fine-tuned for these objects. 
This strategy could also be used for other cases than transparency if specialized foundation models exist.

At the same time, we reduce the reliance on the segmentation model by detecting transparent objects through comparison of properties rendered with photometric opacity and our geometry opacity. 
In these cases the depth rendered from photometric opacity $d_{photo}$ will differ significantly (normally higher depth values) from the one rendered with geometry opacity $d_{geo}$ (see \cref{fig:transparency} for an intuition).
%To reduce the reliance on segmentation model, we learn the transparent object from the difference of rendered depth from photometric alpha and rendered depth from geometric alpha, from the intuition that the rendered depth from photometric alpha is larger on transparent objects. \HZ{Give an illustration}
%\ZL{three sentences were here starting with To improve, To reduce, To reduce, should be avoided}
This is implemented as follows
\begin{equation} M_{diff} = \Vert d_{photo} - d_{geo}\Vert > \epsilon,\label{M_diff}\end{equation} 
and the supervision using $M_{full} = M_{seg}\cup M_{diff}$ becomes
\begin{equation}
    \mathcal{L}_d^{VGGT} = (1-M_{full}) \odot \|d_r - \tilde{d}_{VGGT}\|,\quad \mathcal{L}_d^{DKT} = M_{full} \odot \|d_r - \tilde{d}_{DKT}\|,
\end{equation}
with the notation of $\tilde{d}_{VGGT, DKT}$ defined in \cref{sub:learningincycles}. Note that $\tilde{d}_{VGGT, DKT}$ should first align with $d_r$ before calculating the loss.
% $M_{seg}$ and $M_{diff}$ are used to mask predictions later on. 
% it means that photometric multi-view consistency cannot determine the geometry. 
% \ZL{Explanation what happens in these cases, low confidence? pixel offset? but how? more details needed, from this text its also not clear where the "in cycles" comes from}
% The threshold triggers mostly for transparent or missing parts, or for areas with no feature information, \eg flat plainly-colored parts, and a lack of different viewing angles. 
% \ZL{I propose one figure with one input image, the transparent/not-transparent segmentation, and the thresholded map, and the predictions}

\subsection{Learning in Cycles} \label{sub:learningincycles}
As the output of pretrained depth models is used directly to supervise, the performance of these models is a bottleneck even though we know they do not perform perfectly. 
%As the method relies on the performance of pretrained models, its robustness is bottle-necked by the models. 
To improve robustness, we propose reducing this to a soft reliance by learning a pixel-wise offset map for every depth estimation to rectify the false perception. For each training view, we align them with two learnable offset depth maps $D^{VGGT}$ and $D^{DKT}$ for the visual foundation model (VGGT) and fine-tuned model (DKT), respectively. The offset maps contribute to each $\mathcal{L}_d$ via (written in one equation because they are applied the same way):
\begin{equation}
    \tilde{d}_{VGGT, DKT} = d_{VGGT, DKT} + D^{VGGT,DKT},
\end{equation}
and the loss will update offset maps together with a penalization term
\begin{equation}
    \mathcal{L}_p^{VGGT, DKT} = \lambda_{p}\|D^{VGGT, DKT}\|_2.
\end{equation}
Both $D^{VGGT, DKT}$ are initialized with zeros. 

\subsection{Transparency-aware Multi-view Stereo Loss} \label{sub:transparencyawaremvs}
Even though the direct depth prediction might be inaccurate in many cases, geometry also can be derived from multi-view stereo information.
Previous methods~\cite{chen2024pgsr,kim2025multiview} have shown this to be an effective regularizer for geometry in open-domain settings.
%The perception model may not generalize well in many cases. On the other hand, geometry can be directly derived from multi-view stereo, as pointed out in many previous methods~\cite{chen2024pgsr}~\cite{kim2025multiview}, and it has been shown to be effective in improving the geometry for open-domain images. 
However, stereo depth might be fooled for (semi-)transparent objects because the appearance comes from behind the surface. 

To overcome this, we use the segmentation of transparent parts $M_{seg}$ again (see \cref{sub:detectionguidedsupervision}), and enforce the photometric multi-view stereo loss on non-transparent objects.
%propose to detect the transparent or semi-transparent parts and only enforce the photometric multi-view stereo loss on non-transparent objects. 
% The multi-view geometric consistency loss is proposed in PGSR~\cite{chen2024pgsr}: 
% \begin{equation} 
%     \mathcal{L}_{mvgeom} = \|p_r - H_{nr}H_{rn}p_r\| 
% \end{equation} 
%We use open-vocabulary segmentation models such as SAM3 to detect transparent parts. 
Let $T$ be the set of pixels masked by $M_{seg}\cup M_{diff}$~\cref{M_diff},
% Let $T \subset I$ the set of pixels in image $I$ for which transparent or semi-transparent objects were detected, and $F = \{ p \in I\ |\ \|d_{photo} - d_{geo}\| > \epsilon\}$ the pixels for which learning in cycles in applied, 
then the multi-view \emph{photometric} consistency loss becomes
%Denote the set of pixels of detected transparent or semi-transparent parts by $T$, and the set of $\{p\ |\ \|d_{photo} - d_{geo}\| > \epsilon\}$ as F, 
\begin{equation}
    \mathcal{L}_{mvrgb}^{mask} = \sum_{p_r\in I-T}(1-NCC(I_r(p_r), I_n(H_{rn}p_r)))
\end{equation}
where $NCC$ is the normalized cross correlation of patches~\cite{yoo2009fast}.
The full loss function becomes 
\begin{equation}
\begin{aligned}
    \mathcal{L} =& \mathcal{L}_{rgb} + \mathcal{L}_{d}^{VGGT, DKT} + \mathcal{L}_{n}^{VGGT, DKT} + \mathcal{L}_p^{VGGT, DKT} \\ &+\mathcal{L}_{svgeom} + \mathcal{L}_{mvgeom} + \mathcal{L}_{mvrgb}^{mask}.
\end{aligned}
\end{equation}

\begin{table}[t]
    \centering
    \begin{tabular}{l ccc c ccc} 
    \toprule
         & \multicolumn{3}{c}{\text{NeRF Synthetic}} & & \multicolumn{3}{c}{\text{TransLab}}\\
         \cmidrule{2-4}
         \cmidrule{6-8}
         & PSNR$\uparrow$ & CD$\downarrow$ & F1$\uparrow$ & & PSNR$\uparrow$ & CD$\downarrow$ & F1$\uparrow$  \\ 
         \midrule
     \text{2DGS}~\cite{Huang2DGS2024} & 32.97 & 0.018 &0.325 & &35.68 & 2.355 & 0.917 \\
     \text{+ GT depth}& 30.80 &\cellcolor{red!20}0.011 &\cellcolor{red!20}0.615 & & 35.82 & \cellcolor{red!20}1.373 & \cellcolor{red!20}0.966\\
     \text{+ opacity$_{geo}$}& \cellcolor{orange!20}33.24 & 0.014& 0.410& & \cellcolor{orange!20}37.59& 7.173 & 0.640\\
     \text{+ opacity$_{geo}$ + GT depth $\quad$} & \cellcolor{red!20}33.29 & \cellcolor{orange!20}0.012& \cellcolor{orange!20}0.594 & & \cellcolor{red!20}37.68 & \cellcolor{orange!20}1.542 & \cellcolor{orange!20}0.962\\
     \midrule
     \text{PGSR}~\cite{chen2024pgsr} & 31.76 & 0.016&0.509 & &38.21 & 2.508 & 0.882\\
     \text{+ GT depth}& 29.78 & \cellcolor{red!20}0.010& \cellcolor{red!20}0.729& &35.32 & \cellcolor{orange!20}1.722 & \cellcolor{orange!20}0.961\\
     \text{+ opacity$_{geo}$}& \cellcolor{orange!20}32.29 & 0.021& 0.467& &\cellcolor{red!20}39.37 & 2.745 & 0.881\\
     \text{+ opacity$_{geo}$ + GT depth} & \cellcolor{red!20}33.06 & \cellcolor{red!20}0.010& \cellcolor{orange!20}0.706& &\cellcolor{orange!20}39.34 & \cellcolor{red!20}1.653 & \cellcolor{red!20}0.964\\
     \bottomrule
    \end{tabular}
    \caption{The effect of using ground-truth depth with and without geometry opacity. Here $+opacity_{geo}$ is simply a parameter addition for rendering geometry. In general, using GT depth without geometry opacity improves geometric accuracy but hurt the rendering. Using both together enables outstanding results for both rendering and geometry. F1 score is much lower than 1 even with the best reconstruction because the NeRF-Synthetic contains invisible parts. (best result in red, second best in orange) }
    \label{exp:quantitative:representation power}
\end{table}

\begin{figure}[t]
    \centering
    \begin{tabular}{ccccc}
      \begin{overpic}[width=0.18\textwidth]{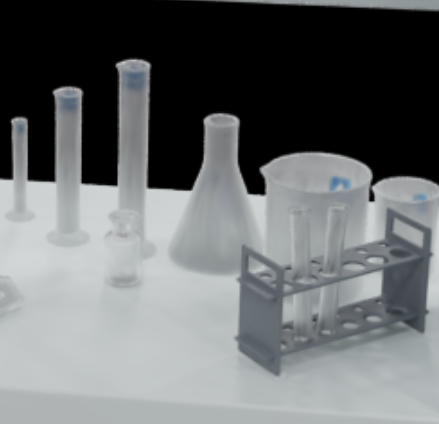} \put(50,50){\color{red}\circle{20}} \end{overpic}  
      & \begin{overpic}[width=0.18\textwidth]{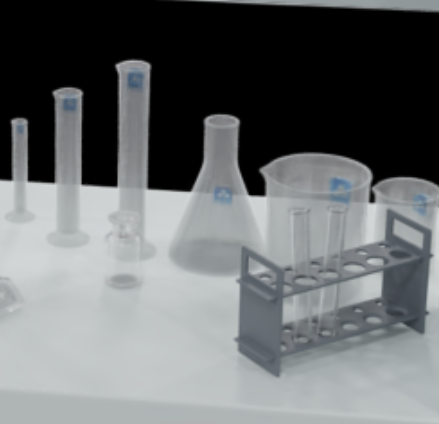} \put(50,50){\color{red}\circle{20}} \end{overpic} 
      & \begin{overpic}[width=0.18\textwidth]{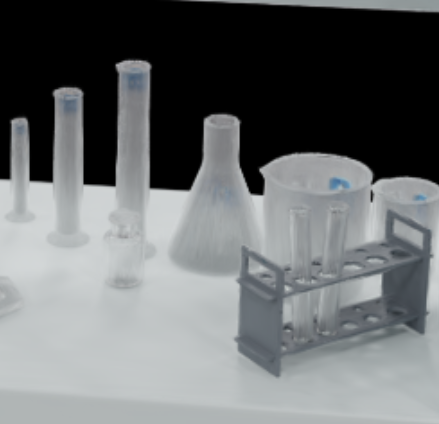} \put(50,50){\color{red}\circle{20}} \end{overpic}
      & \begin{overpic}[width=0.18\textwidth]{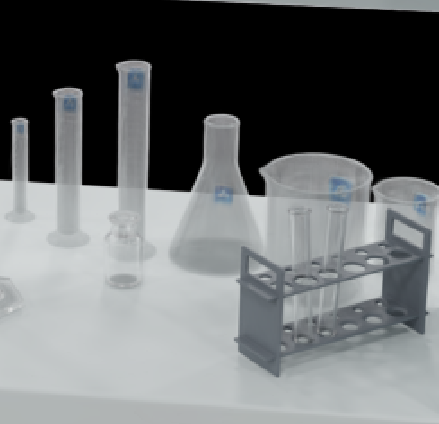} \put(50,50){\color{red}\circle{20}} \end{overpic} 
      & \begin{overpic}[width=0.18\textwidth]{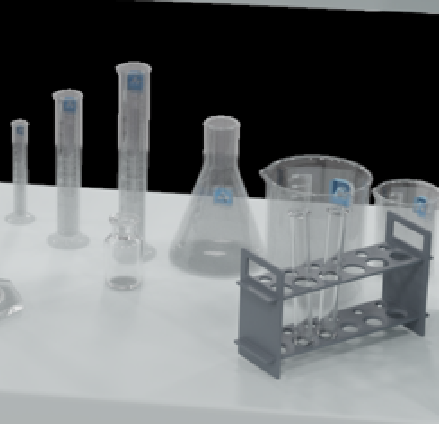}  \put(50,50){\color{red}\circle{20}} \end{overpic} \\
        \makecell[tc]{\textbf{2DGS} \\+ GT depth} & \makecell[tc]{\textbf{2DGS}\\+ GT depth\\+ opacity$_{geo}$} & \makecell[tc]{\textbf{PGSR} \\+ GT depth} & \makecell[tc]{\textbf{PGSR}\\+ GT depth \\+ opacity$_{geo}$} & \makecell[tc]{\textbf{GT}}
    \end{tabular}
    \caption{Rendering result for experiments with ground-truth depth and geometry opacity. Without the separation of geometry and rendering through opacity$_{geo}$ the rendering removes details on transparent objects. }
    \label{exp:qualitative:representation power}
\end{figure}

\section{Experiments} \label{sec:experiments}

This section contains all experimental results. Details about the implementation can be found in the supplementary material. 

\subsection{Datasets}
%We evaluate the rendering quality on Mip-NeRF360~\cite{barron2022mip}, TransLab~\cite{li2025tsgs} and NeRF Synthetic~\cite{mildenhall2021nerf} and we evaluate the reconstruction quality on TransLab, DTU~\cite{jensen2014large} and NeRF Synthetic.
We use the following datasets for evaluation:

\noindent\textbf{TransLab}~\cite{li2025tsgs}. Eight scenes from a lab environment with many transparent objects and ground-truth geometry.

\noindent\textbf{NeRF Synthetic}~\cite{mildenhall2021nerf}. Eight synthetically rendered scenes with complex non-Lambertian material and ground-truth geometry.

\noindent\textbf{Mip-NeRF360}~\cite{barron2022mip}. Nine complex indoor and outdoor scenes, for rendering only. 

\noindent\textbf{DTU}~\cite{jensen2014large}. 80 everyday objects with ground-truth geometry but no setup for comparison of rendering.

\subsection{Baseline models}

\noindent\textbf{2DGS}~\cite{Huang2DGS2024}.  Restricts geometry by restricting the Gaussian covariance to be two-dimensional. We base the experiments on the scripts of the original 2DGS implementation for Mip-NeRF360, DTU and NeRF Synthetic. We apply the parameters of DTU to the ones of TransLab. 

\noindent\textbf{PGSR}~\cite{chen2024pgsr}. Regularizes the surface properties by penalizing non-flat splats. We base the experiments on the scripts as the original PGSR implementation for Mip-NeRF360 and DTU. We apply the parameters of DTU to the ones of TransLab and NeRF Synthetic.
% To train on TransLab, we use the parameters of DTU. To train on NeRF Synthetic, we use the parameters of MipNeRF-360.

\noindent\textbf{GOF}~\cite{yu2024gaussian}. Jointly optimizes a mesh and encourages splats to lie on the surface. We compare with it on DTU and MipNeRF-360 as it is not suited for complex materials.

\noindent\textbf{TSGS}~\cite{li2025tsgs}. SOTA method specifically for the reconstruction of transparent objects. Used on TransLab only since it is time-consuming and memory-consuming, and scripts for other scenes are not well provided.

\noindent\textbf{CarGS}~\cite{shen2025evolving}. Recent work for decoupling geometry from appearance through separate covariance matrices. Since the code is not available, we only compare on MipNeRF-360 and copy the numbers from their paper. 

\begin{table}[t]
    \centering
    \begin{tabular}{l ccc c cccc} 
    \toprule
         & \multicolumn{2}{c}{\text{NeRF Synthetic}} & & \multicolumn{3}{c}{\text{TransLab}}\\
         \cmidrule{2-4}
         \cmidrule{6-9}
        & PSNR$\uparrow$ & CD$\downarrow$ & F1$\uparrow$ & & PSNR$\uparrow$ & CD$\downarrow$ & F1$\uparrow$ & Time$\downarrow$\\ 
         \midrule
     \text{2DGS}~\cite{Huang2DGS2024} & \cellcolor{red!20}32.97 & 0.018 & 0.325  & & 35.68 & 2.355 & 0.917 & 0.67h \\
     \text{PGSR}~\cite{chen2024pgsr} & 31.76 & \cellcolor{orange!20}0.016 & \cellcolor{red!20}0.509 & & \cellcolor{orange!20}38.21& 2.508& 0.882 &\cellcolor{red!20}0.5h\\
     \text{TSGS}~\cite{li2025tsgs} & {\centering /} &{\centering /}  & {\centering /} & & 37.39& \cellcolor{orange!20}1.870 & \cellcolor{orange!20}0.954 & 1.5h\\
     \text{TSGS}~\cite{li2025tsgs}$^{\star}$ & {\centering /} &{\centering /}  & {\centering /} & & 39.08& 1.850 & 0.950 & 1.5h\\
     \text{Ours} & \cellcolor{orange!20}32.52 & \cellcolor{red!20}0.015 & \cellcolor{orange!20}0.503 & & \cellcolor{red!20}39.95 &\cellcolor{red!20}1.665 &\cellcolor{red!20} 0.960& \cellcolor{red!20}0.5h \\
     \bottomrule
    \end{tabular}
    \caption{Comparisons in rendering and reconstruction on datasets with highly reflective (NeRF Synthetic) and transparent (TransLab) objects. Our method excels especially on data with transparency. TSGS is not suited for reflective material. ${\star}$ results are copied from the original paper. (best result in red, second best in orange)}
    \label{exp:quantitative:translab}
\end{table}

\subsection{With Ground-truth Geometry}\label{subsec:rep}
Experiments corresponding to \cref{sec:geometricrepresentation} where dense ground-truth information of geometry is provided during the optimization of Gaussian Splatting to test the limit of representational power with one single opacity value and with our additional geometry opacity. 
We tested the differences using 2DGS and PGSR, both popular geometry-aware GS methods, over two synthetic datasets TransLab and NeRF Synthetic which provide ground-truth geometry. 
We report quantitative results in \cref{exp:quantitative:representation power} and, even with complete geometric information, the novel-view rendering is bottle-necked. 
However, adding the geometry opacity unleashes the representation power. 
As shown in \cref{exp:qualitative:representation power}, the label behind the glass becomes visible after adding the geometry opacity.

\begin{figure}[tbh]
    \centering
    \begin{tabular}{cccccc}
        & \textbf{2DGS} & \textbf{PGSR} & \textbf{TSGS} & \textbf{Ours} & \textbf{GT} \\
        \rotatebox{90}{\hspace{4pt} Geometry} & \includegraphics[width=0.18\textwidth]{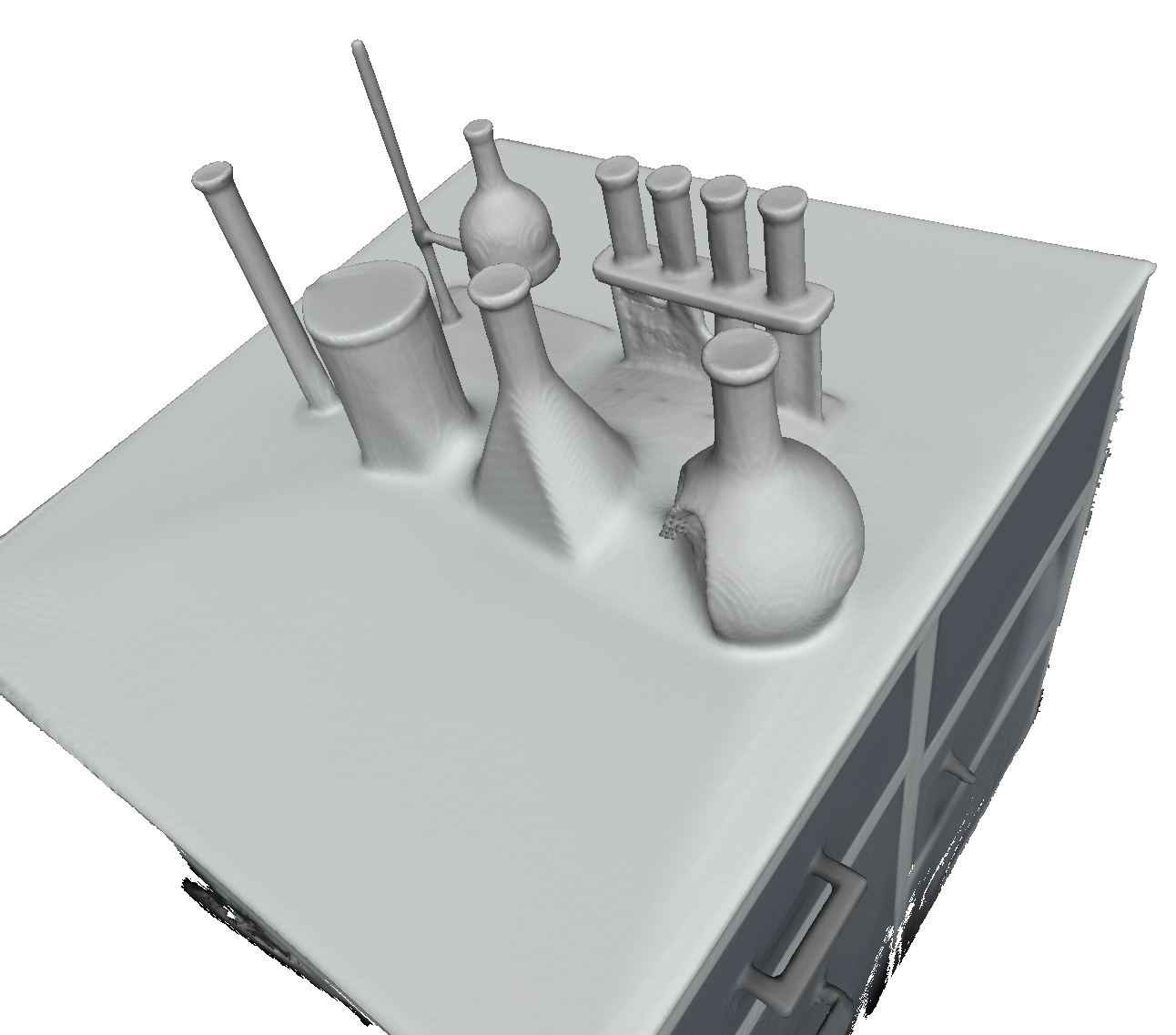} & \includegraphics[width=0.18\textwidth]{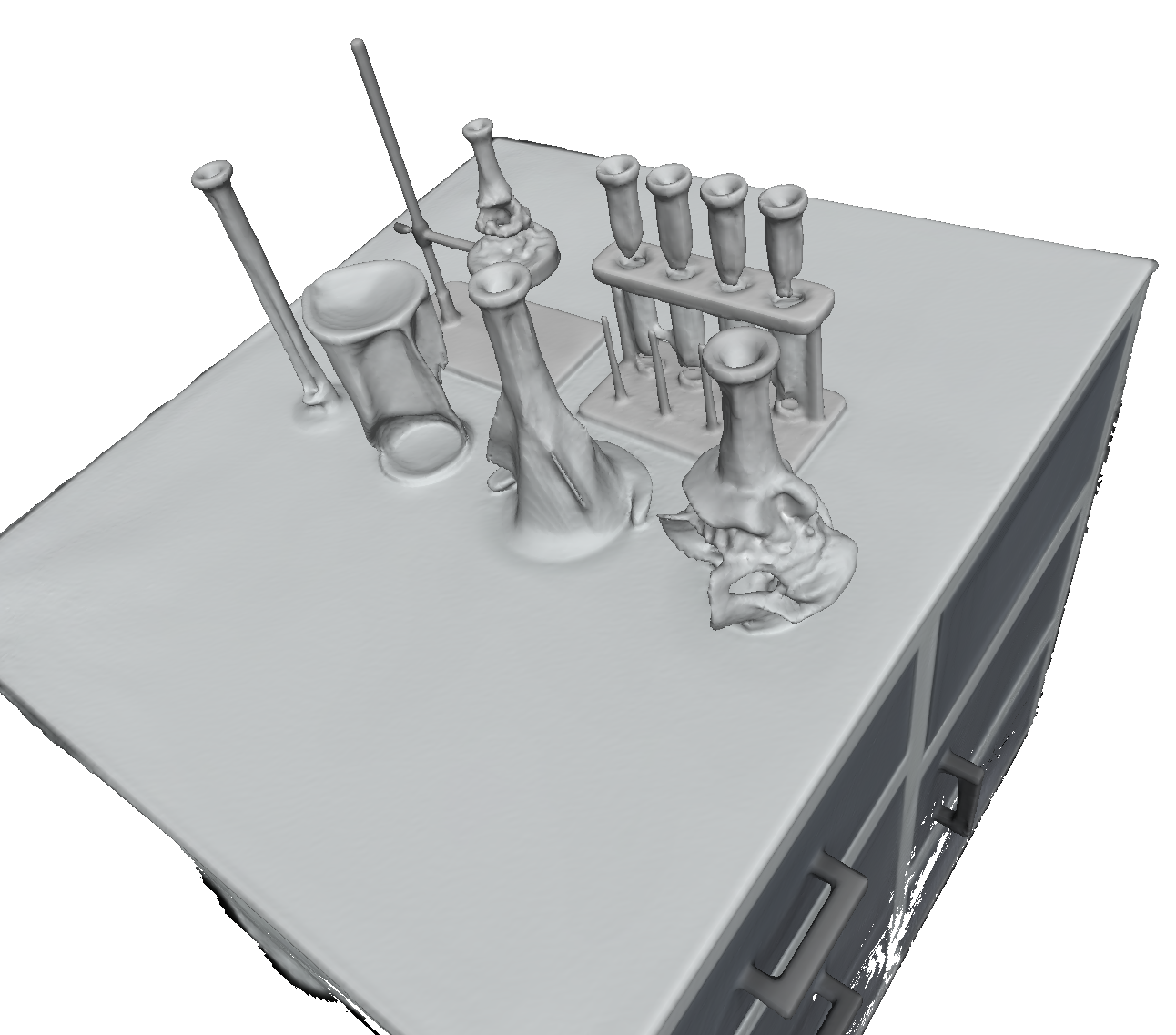} & \includegraphics[width=0.18\textwidth]{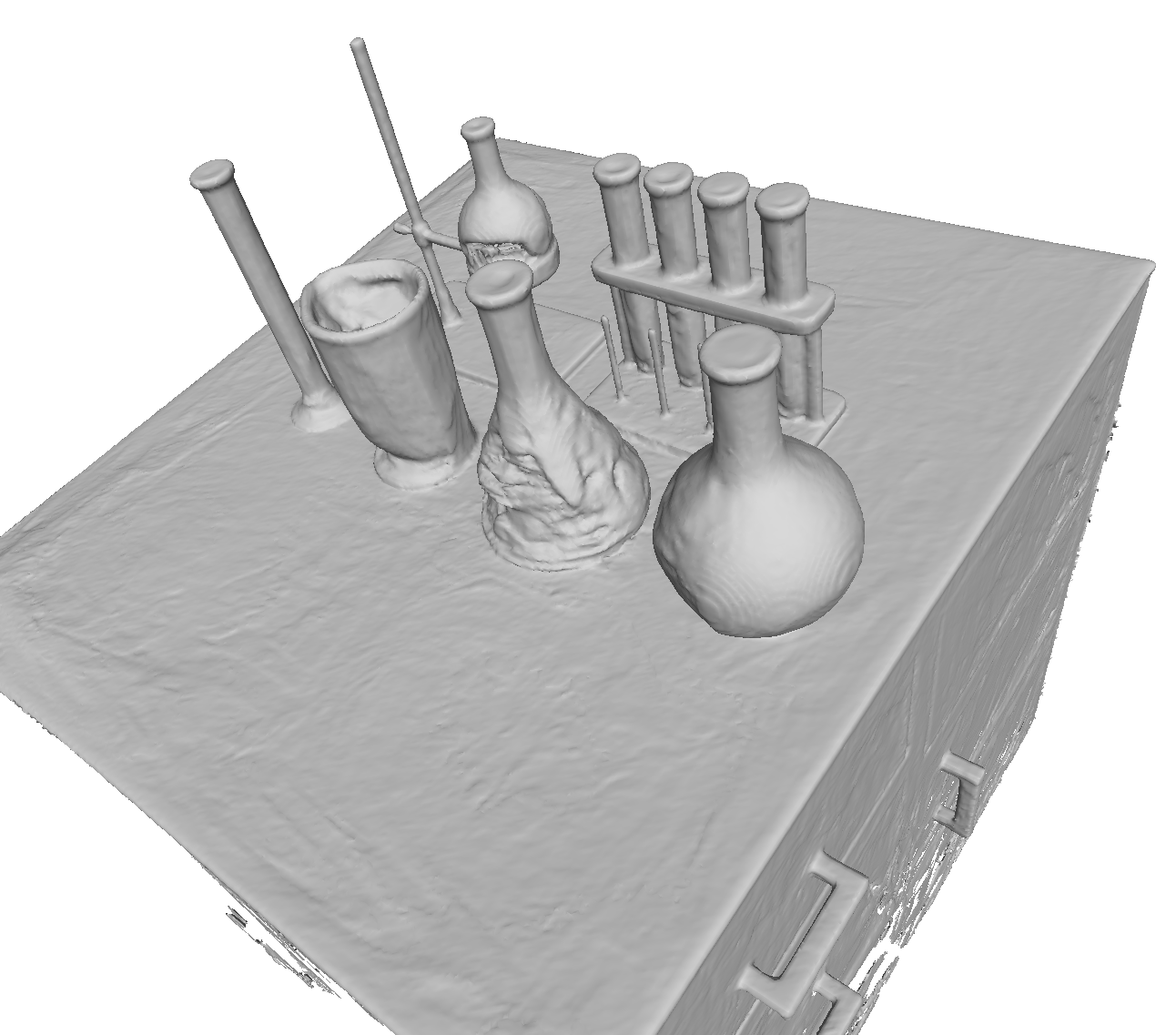} & \includegraphics[width=0.18\textwidth]{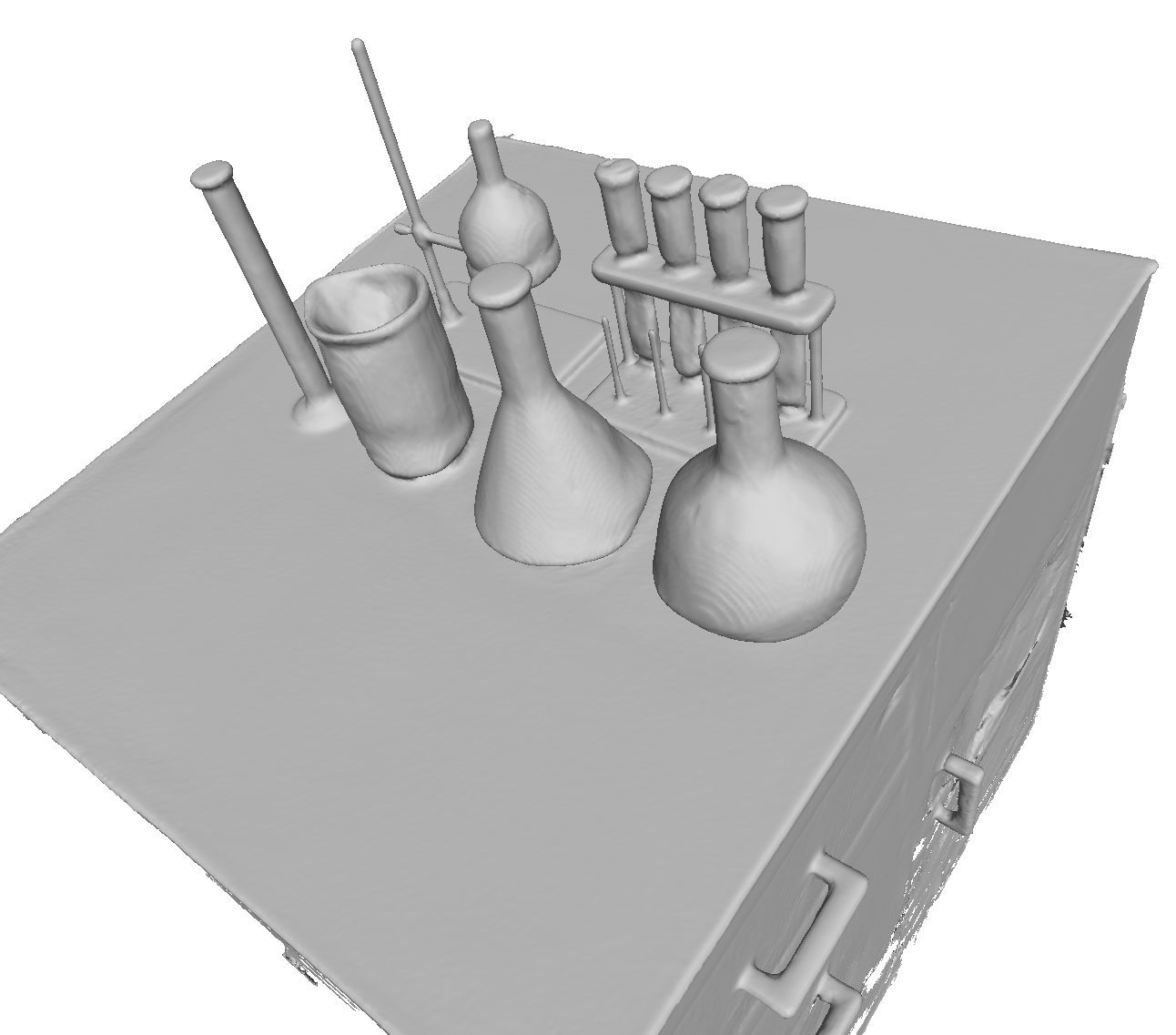} & \includegraphics[width=0.18\textwidth]{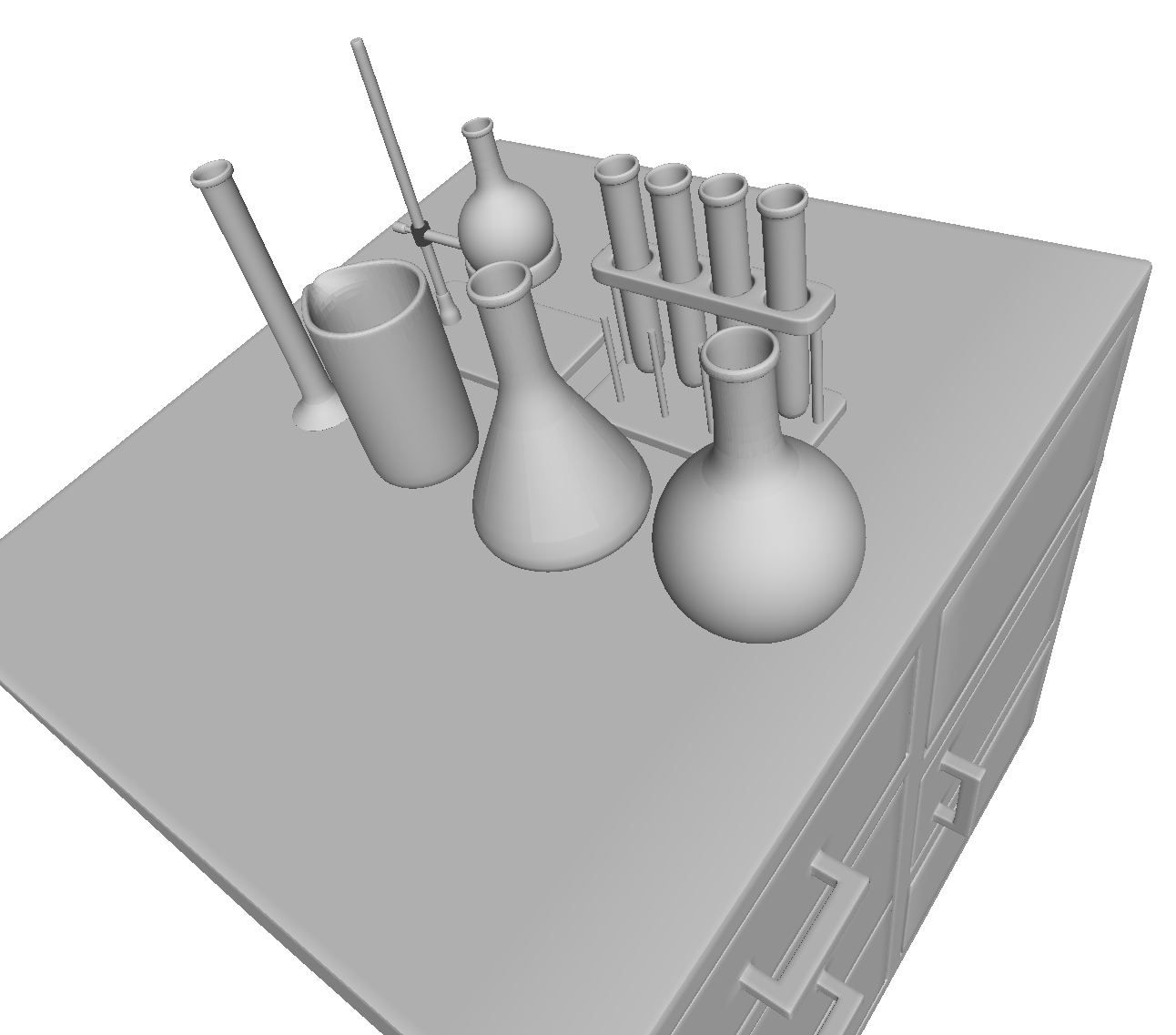} \\
        \rotatebox{90}{\hspace{6pt} Rendering} & \includegraphics[width=0.18\textwidth]{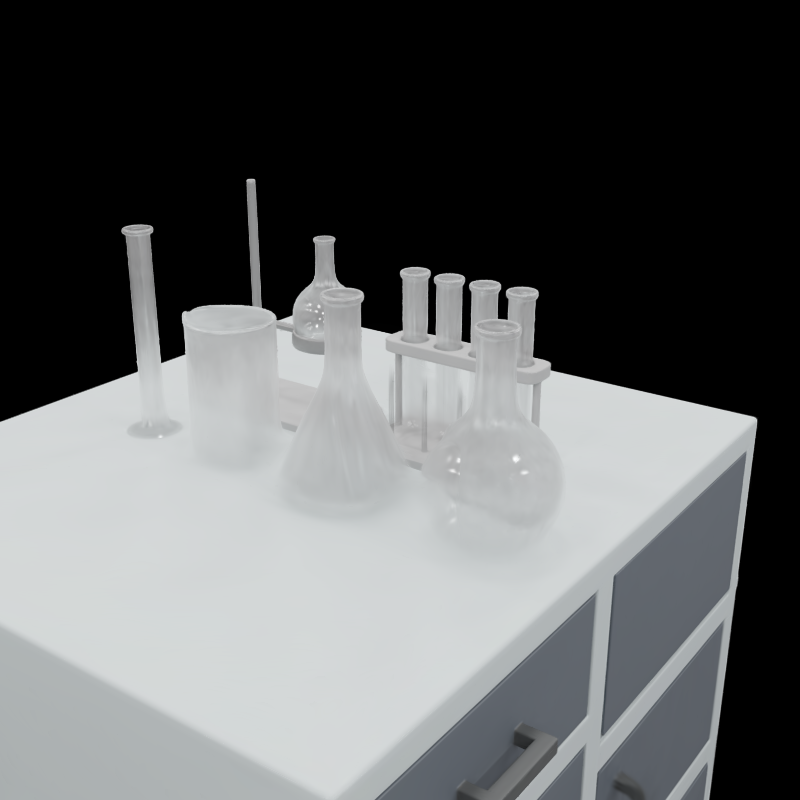} & \includegraphics[width=0.18\textwidth]{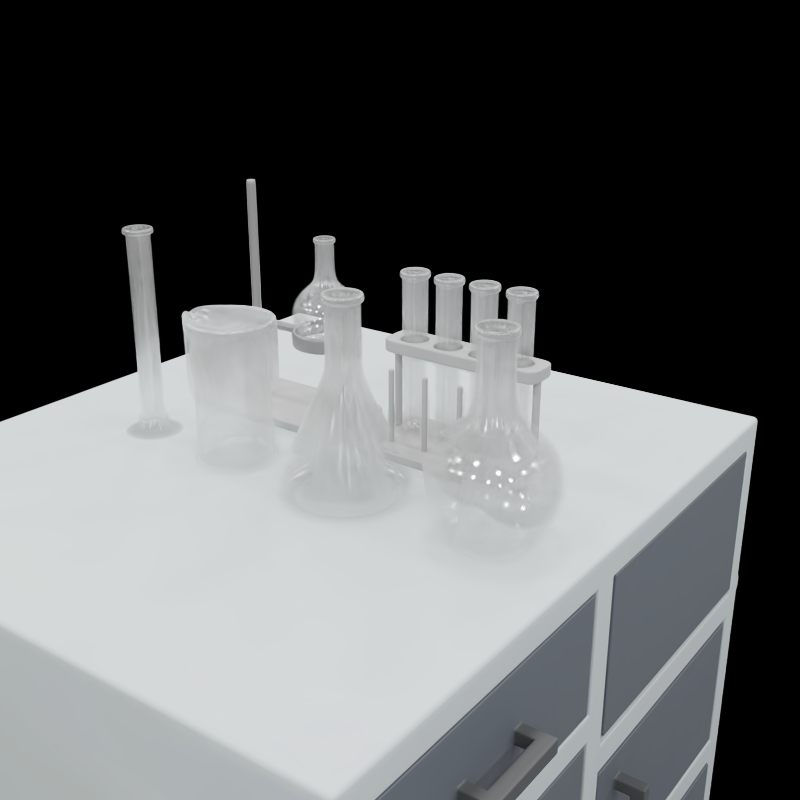} & \includegraphics[width=0.18\textwidth]{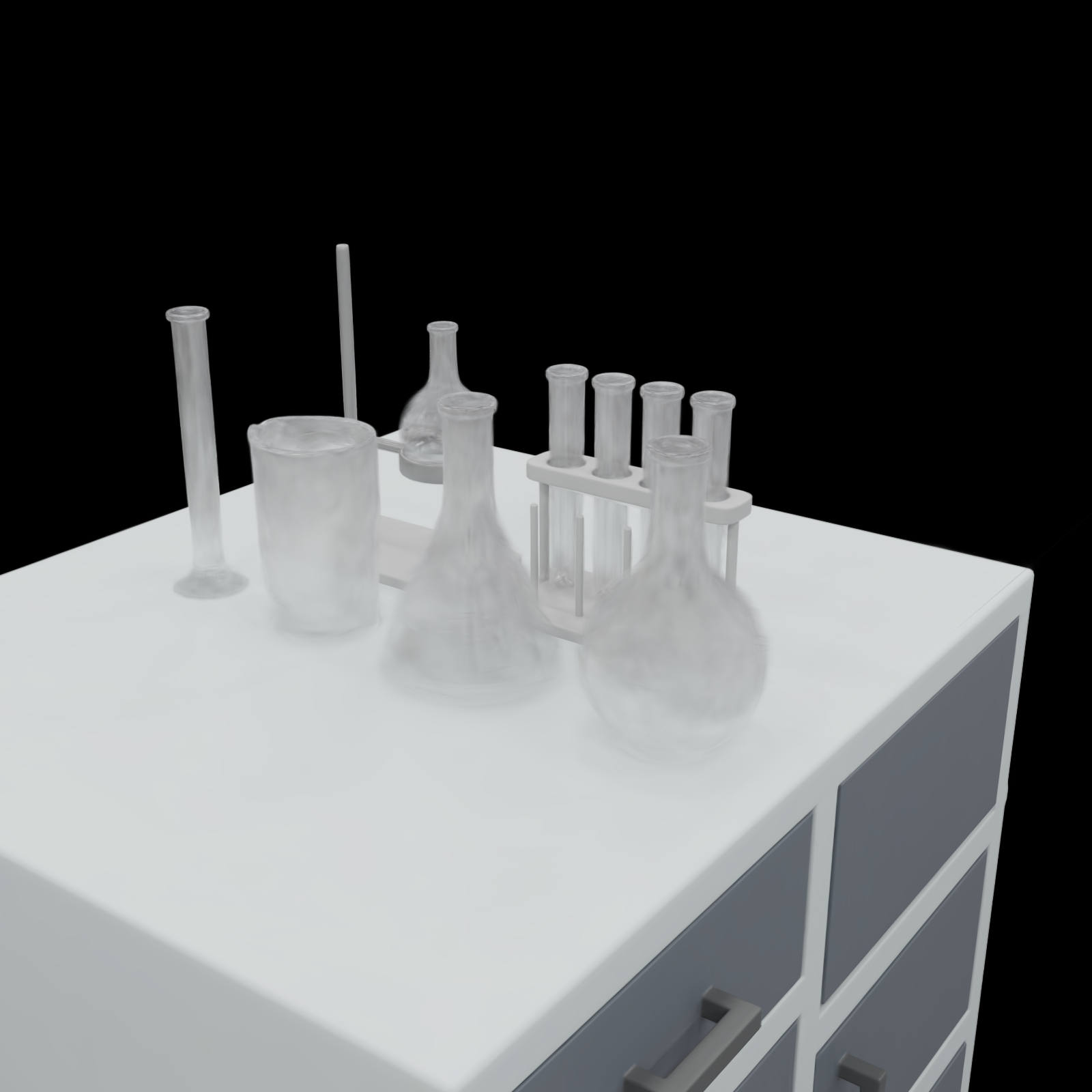} & \includegraphics[width=0.18\textwidth]{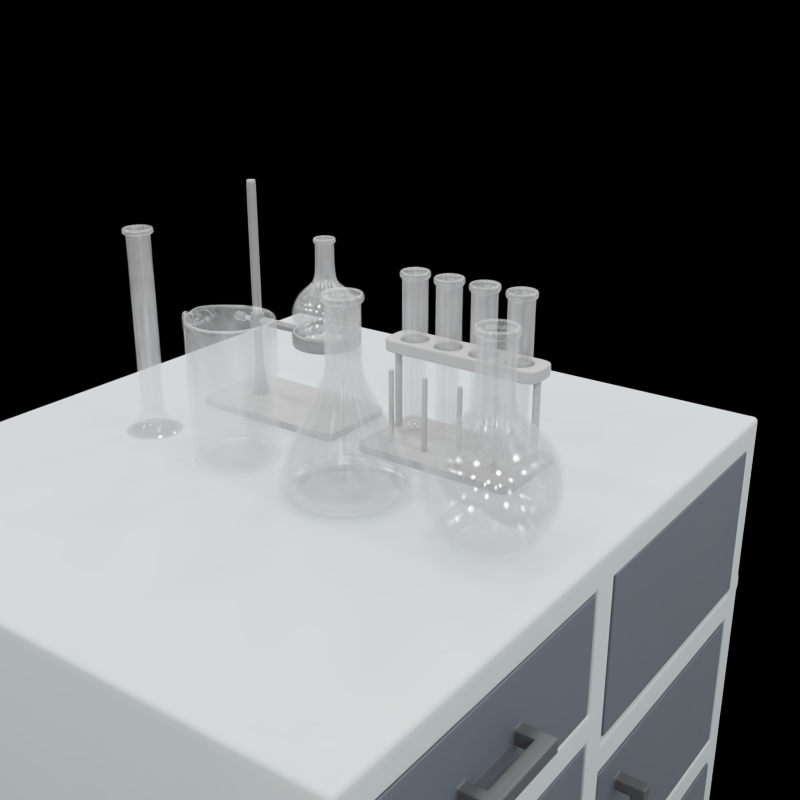} & \includegraphics[width=0.18\textwidth]{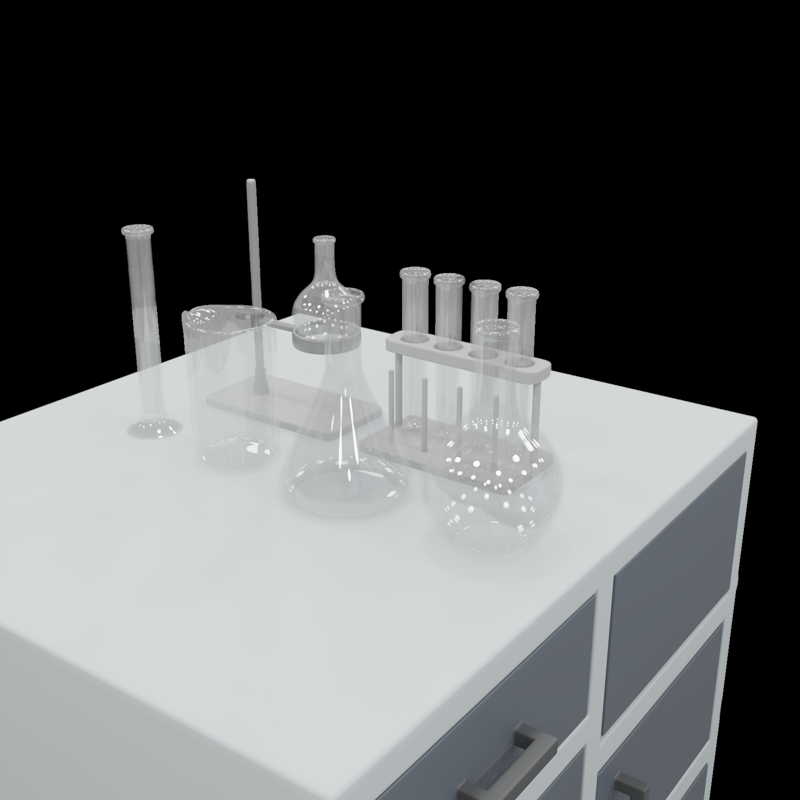} \\
        
        \rotatebox{90}{\hspace{4pt} Geometry} & \includegraphics[width=0.18\textwidth]{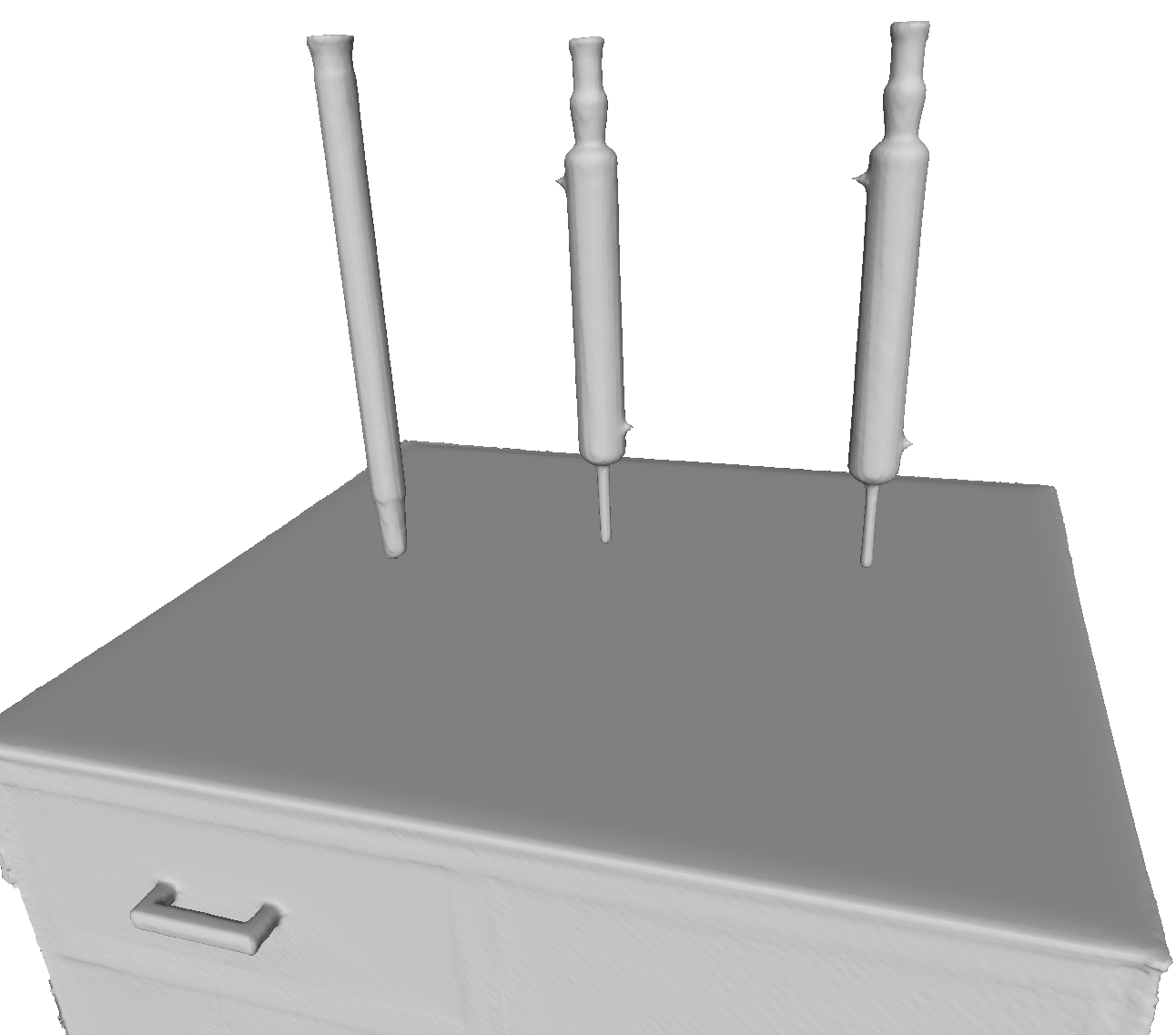} & \includegraphics[width=0.18\textwidth]{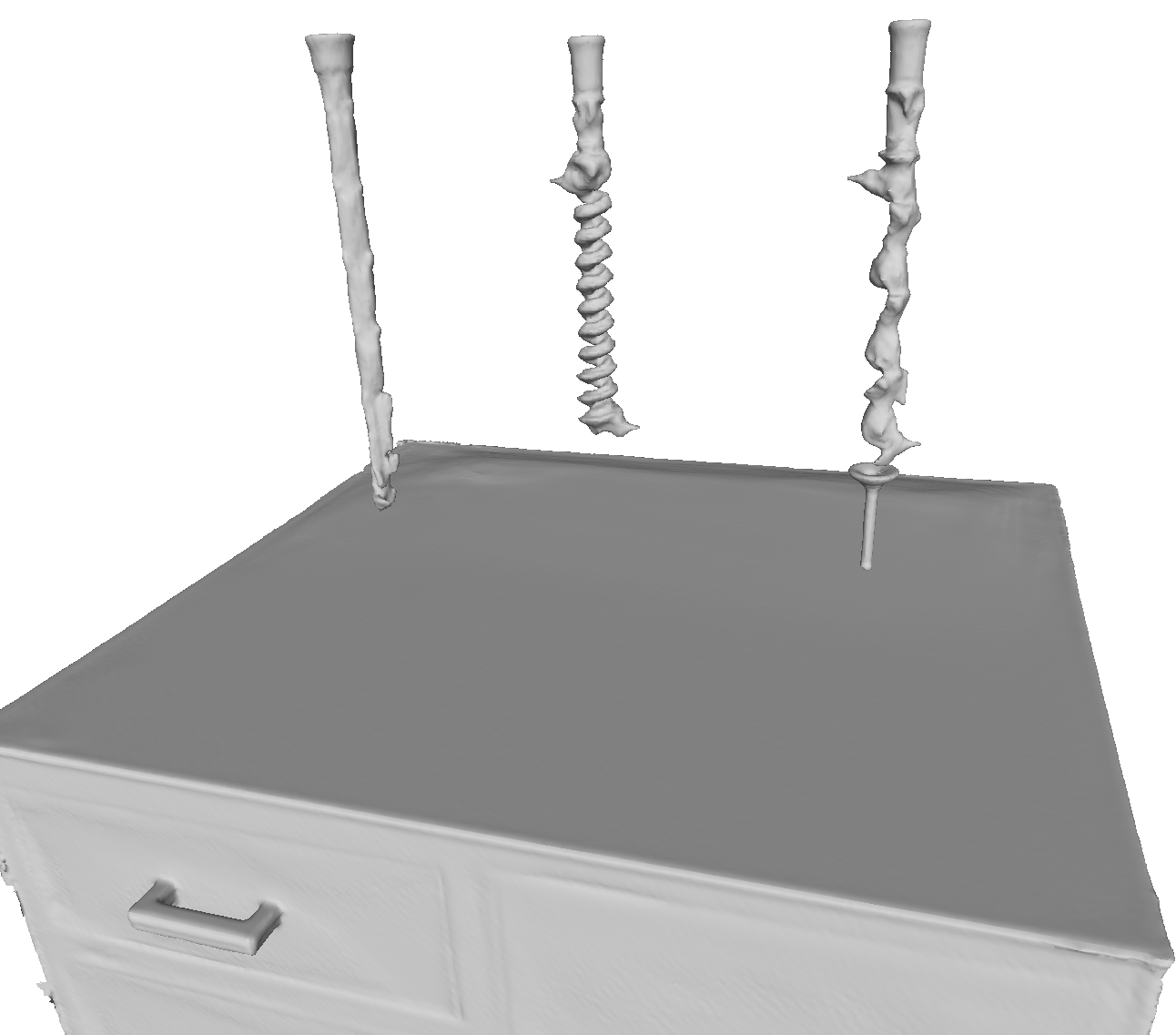} & 
        \includegraphics[width=0.18\textwidth]{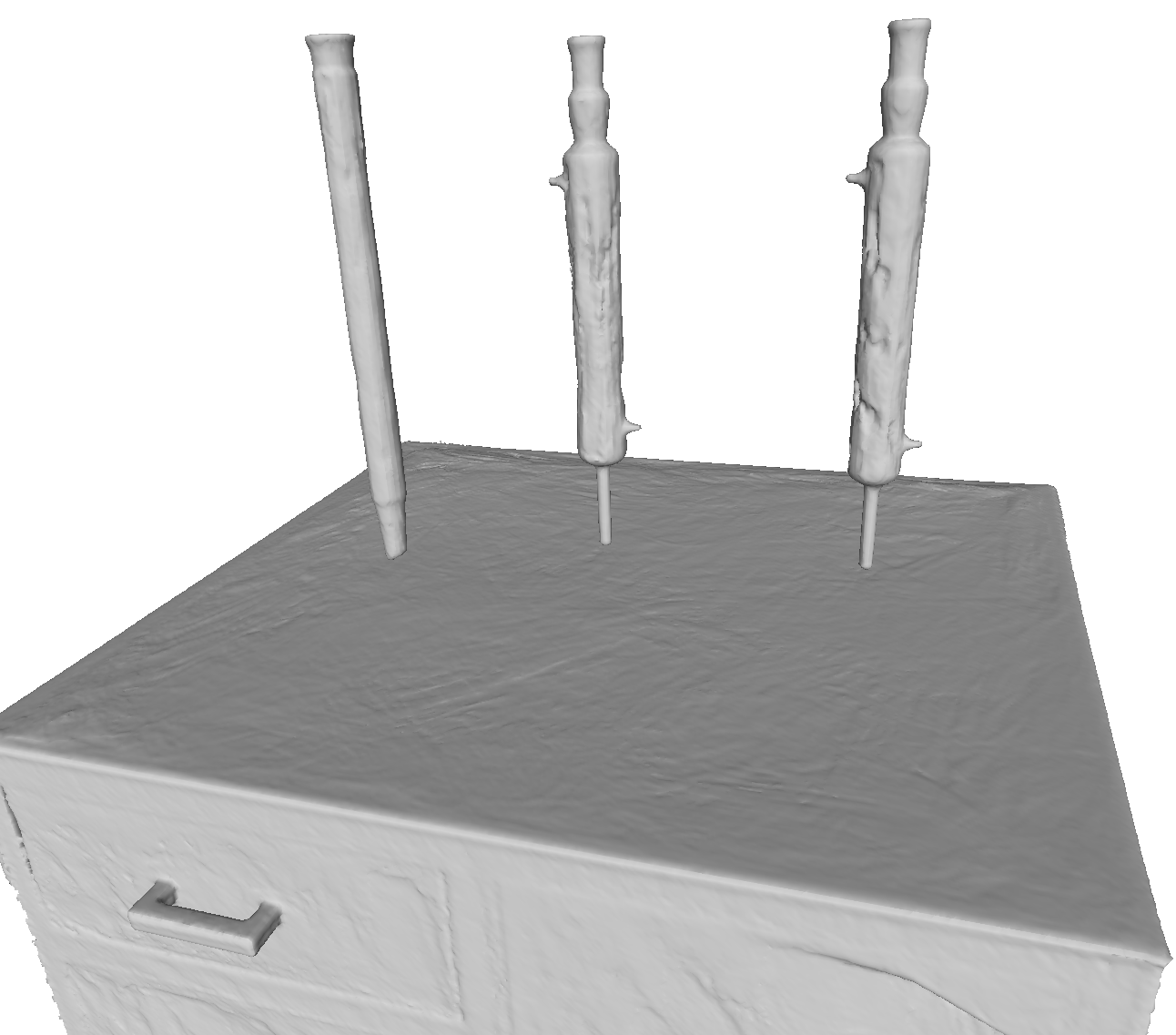} & \includegraphics[width=0.18\textwidth]{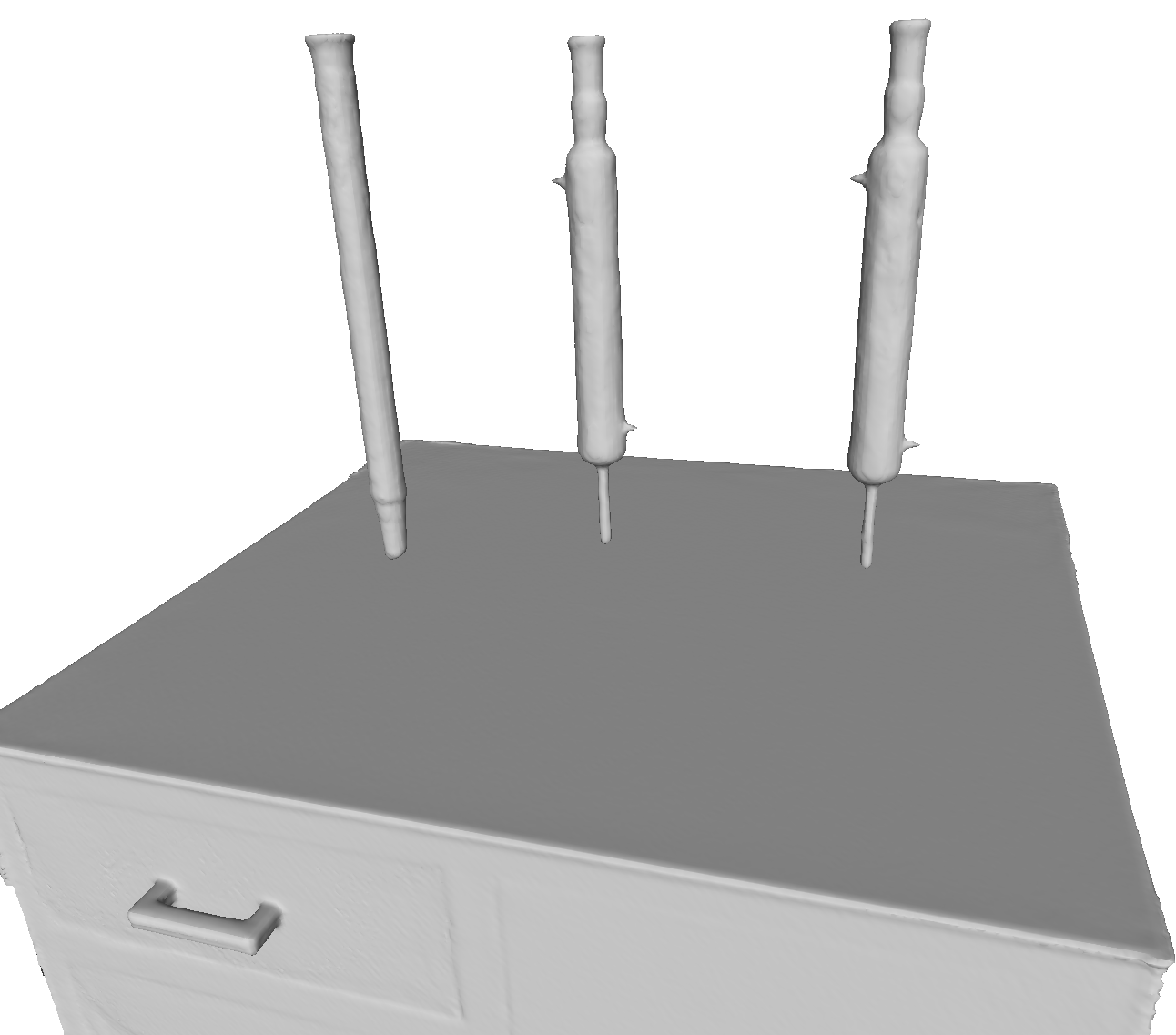} & \includegraphics[width=0.18\textwidth]{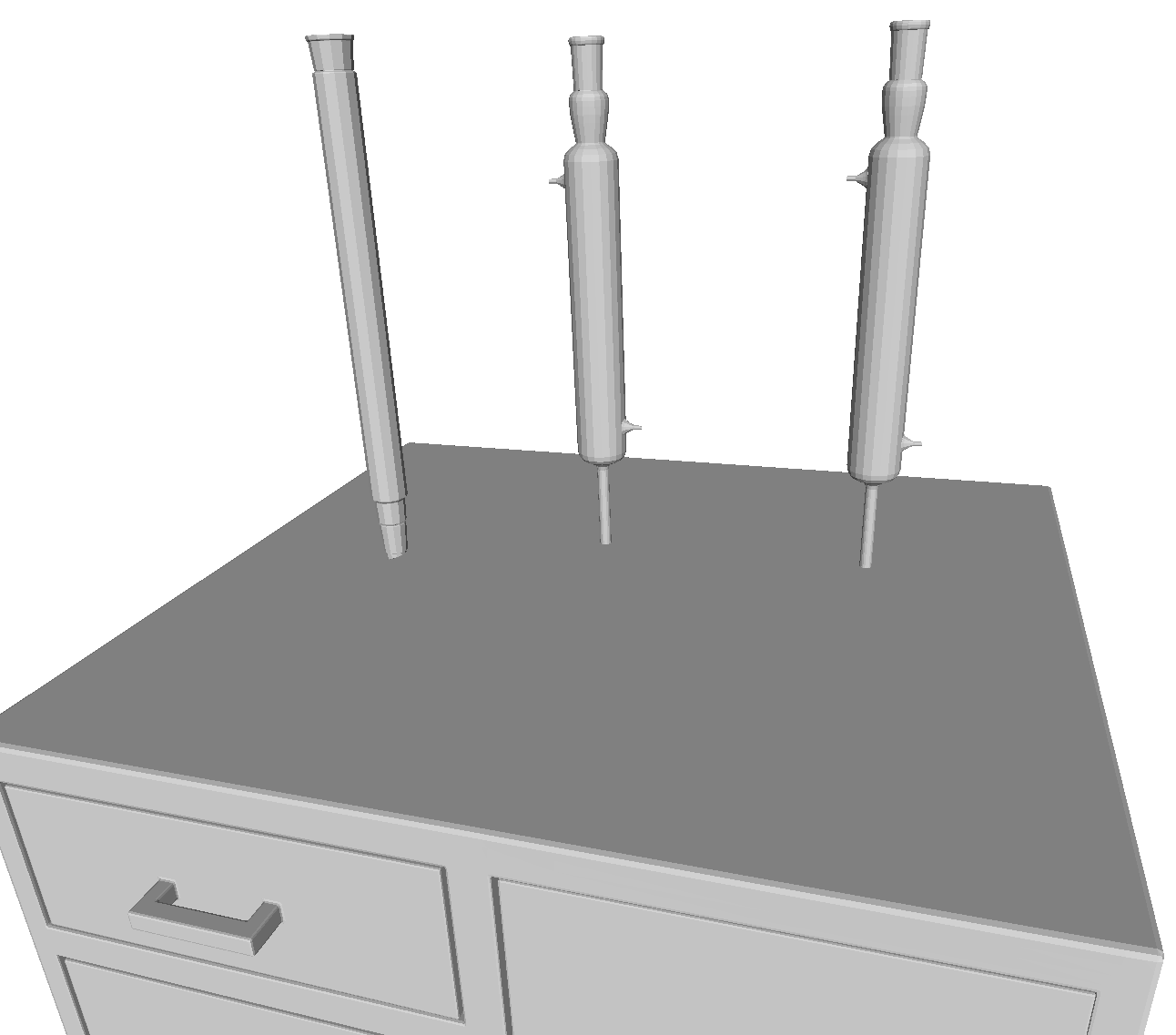} \\
        \rotatebox{90}{\hspace{6pt} Rendering} & \includegraphics[width=0.18\textwidth]{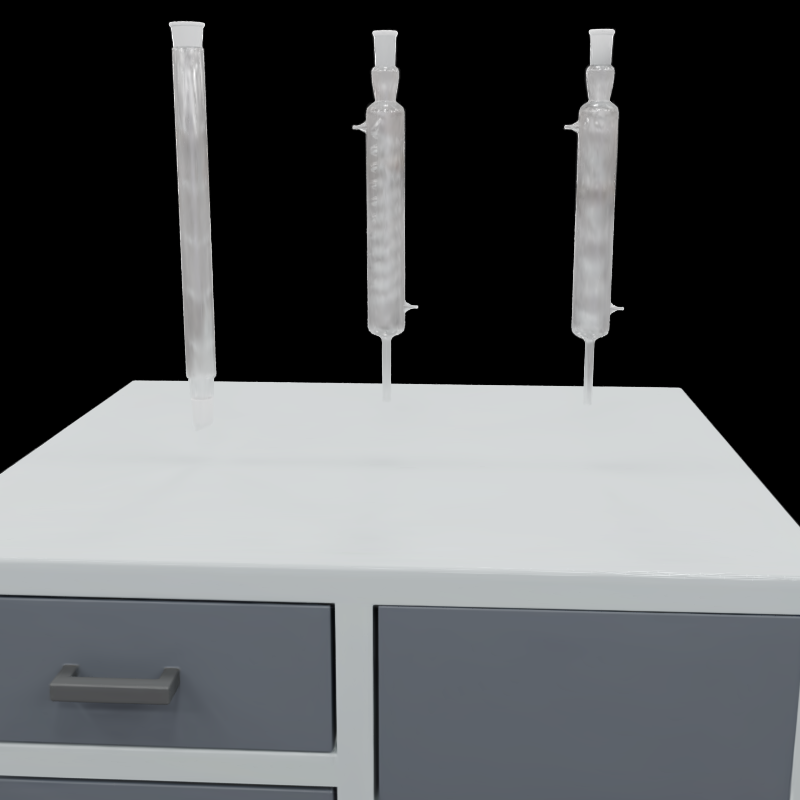} & \includegraphics[width=0.18\textwidth]{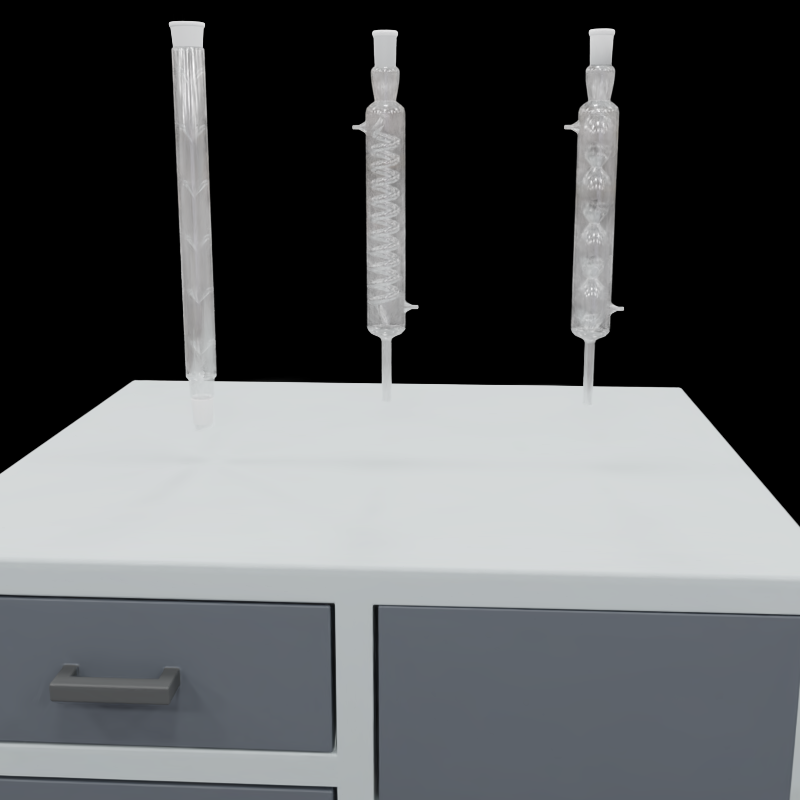} & \includegraphics[width=0.18\textwidth]{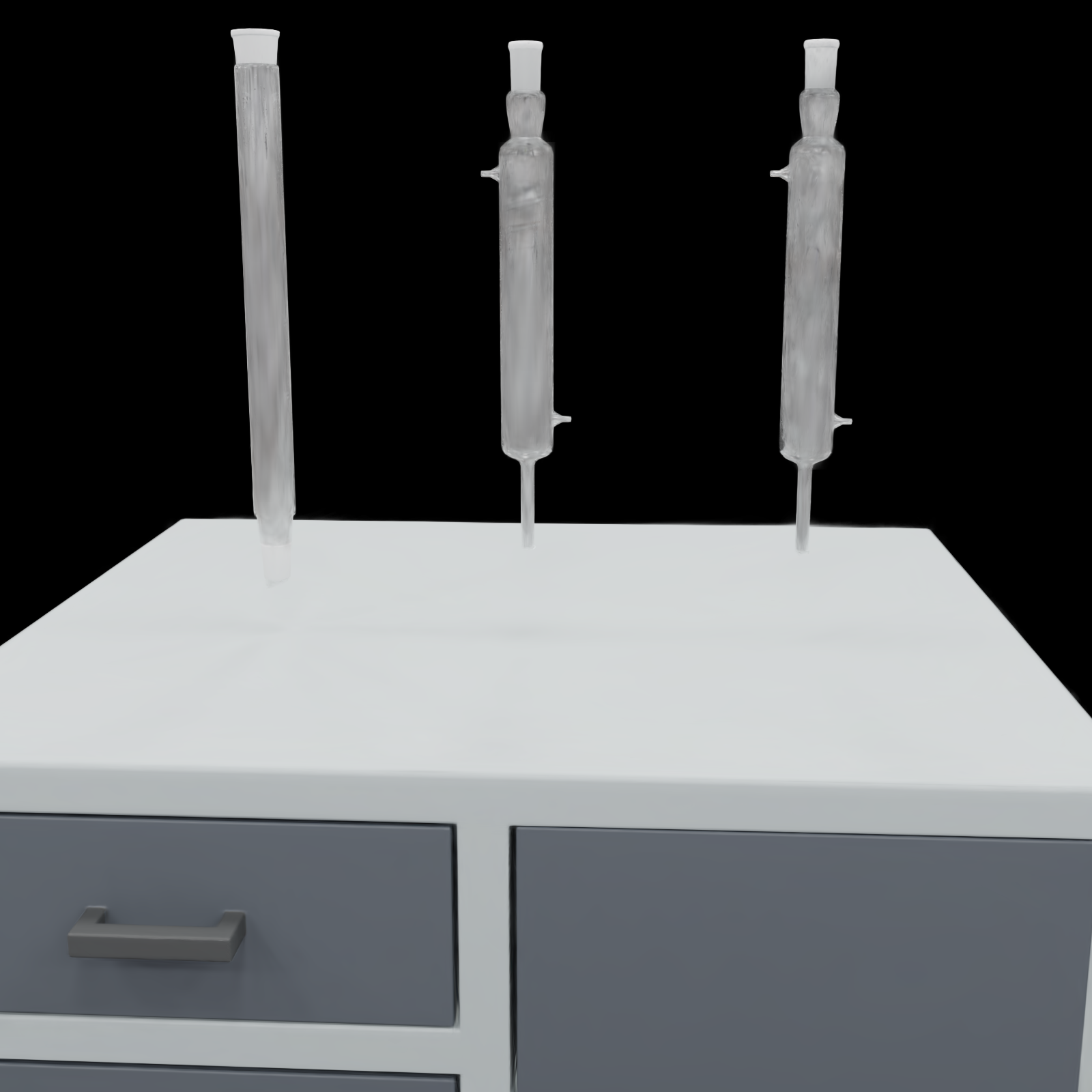} & \includegraphics[width=0.18\textwidth]{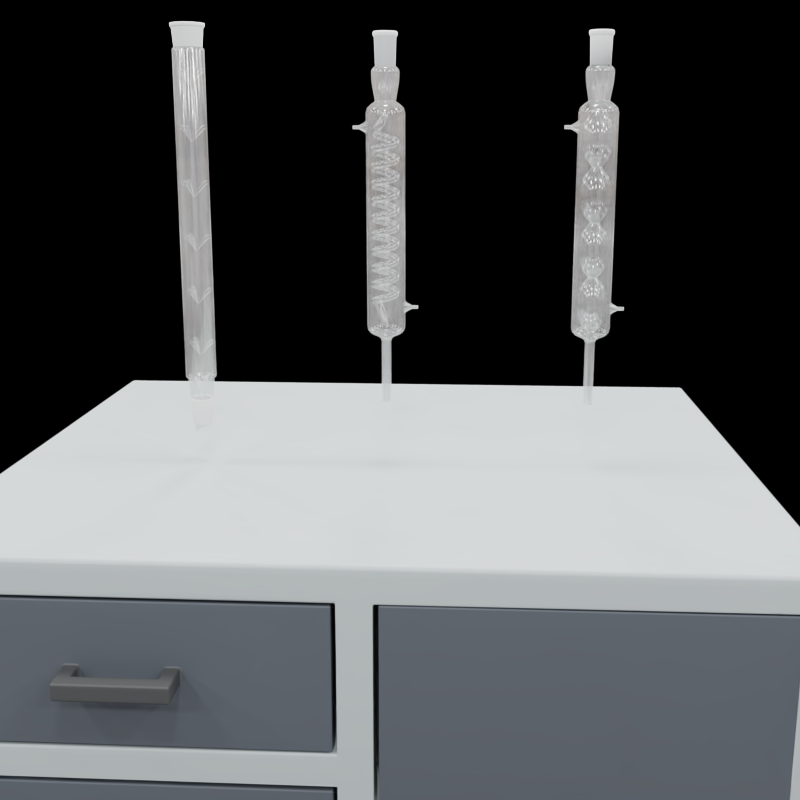} & \includegraphics[width=0.18\textwidth]{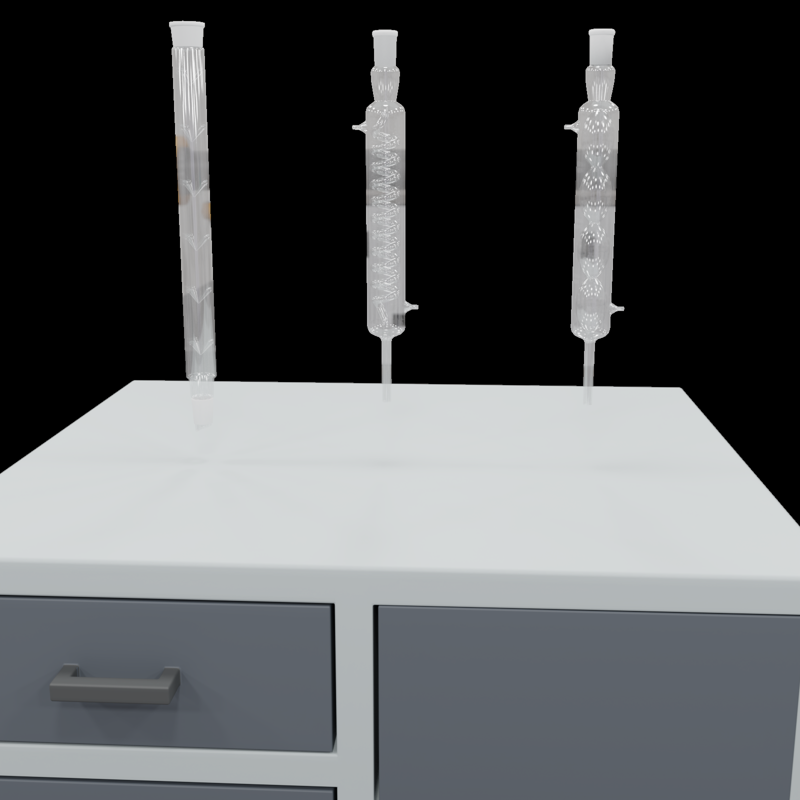} \\

    \end{tabular}
    \caption{Qualitative results on the TransLab dataset. Our results are the most faithful to the ground-truth, even in comparison to transparent-oriented method TSGS, and in presence of highly complex transparent overlaps. }
    \label{exp:qualitative:translab}
\end{figure}

\subsection{Geometry from Vision Foundation Models}
%We can see in \cref{exp:qualitative:representation power} that as long as the geometry information is completely and correctly provided, the Gaussian Splatting is able to attain good geometry and rendering at the same time. 
When using predicted depth from foundation models, small inaccuracies and misalignment is common, especially for transparent surfaces.
In this section, we show results using the pipeline introduced in \cref{sec:method} which combines predicted depth with detection-guided supervision using multi-view stereo information and properties of transparent objects. 
%In this aspect, we propose to combine the geometry derived from multi-view stereo with the geometry supervision from large perception models. 
We evaluate our framework on multiple datasets; covering transparent objects in TransLab, objects with light effect in NeRF-Synthetic and objects with ordinary material in DTU and MipNeRF-360. 

\begin{figure}[tbh]
    \centering
    \begin{tabular}{ccccc}
        & \textbf{2DGS} & \textbf{PGSR} & \textbf{Ours} & \textbf{GT} \\
        \rotatebox{90}{\hspace{13pt} Geometry} & \includegraphics[width=0.23\textwidth]{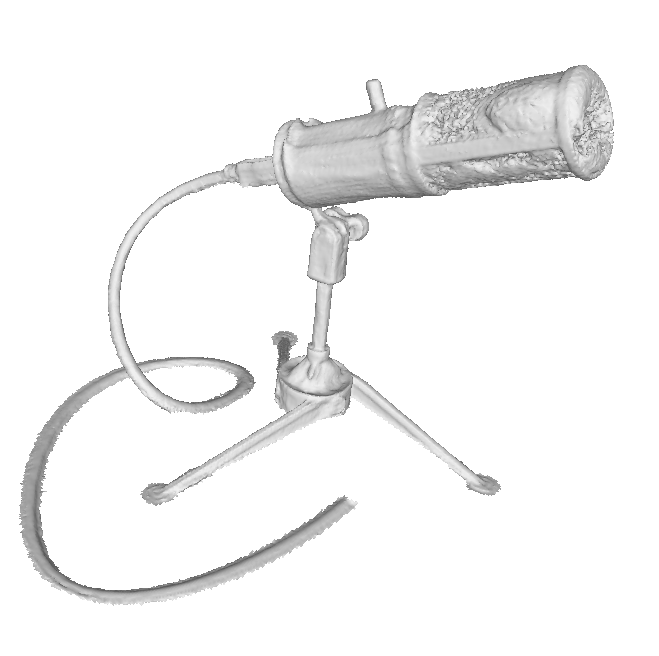} & \includegraphics[width=0.23\textwidth]{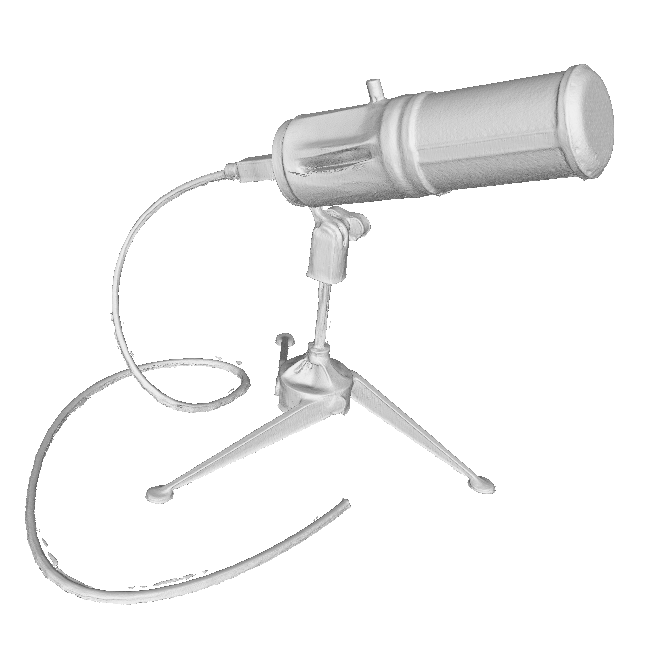} & \includegraphics[width=0.23\textwidth]{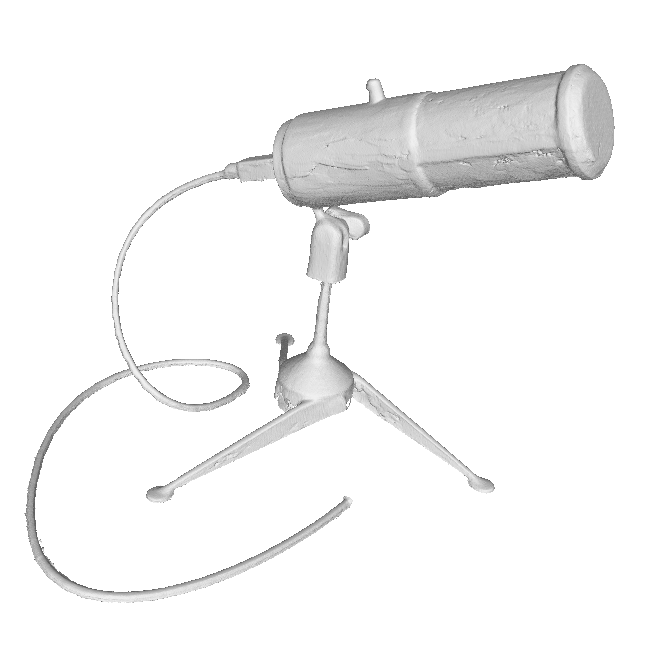} & \includegraphics[width=0.23\textwidth]{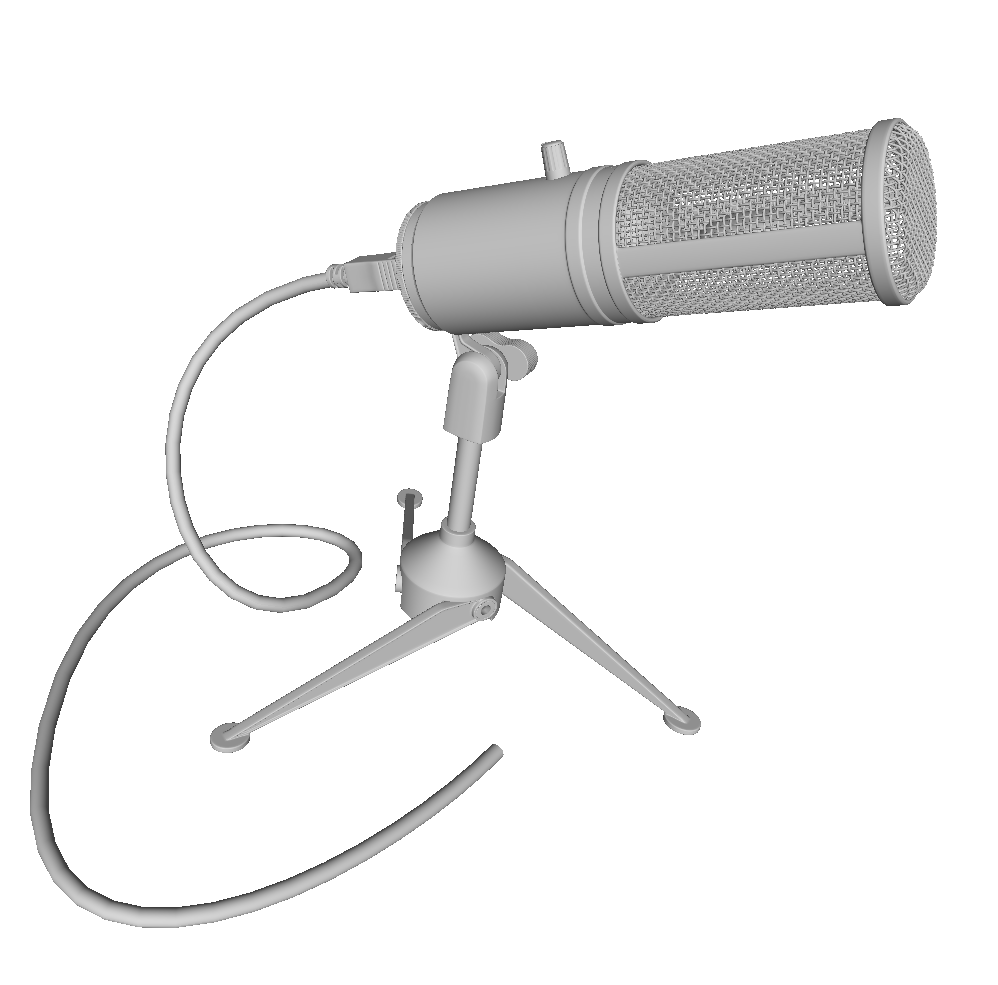}\\
        \rotatebox{90}{\hspace{11pt} Rendering} & \includegraphics[width=0.23\textwidth]{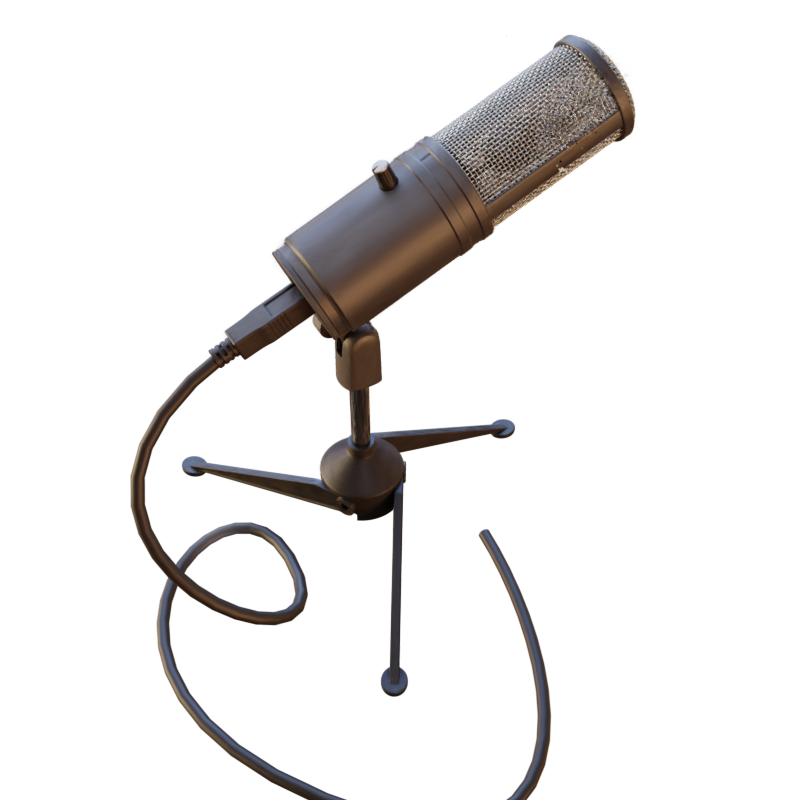} & \includegraphics[width=0.23\textwidth]{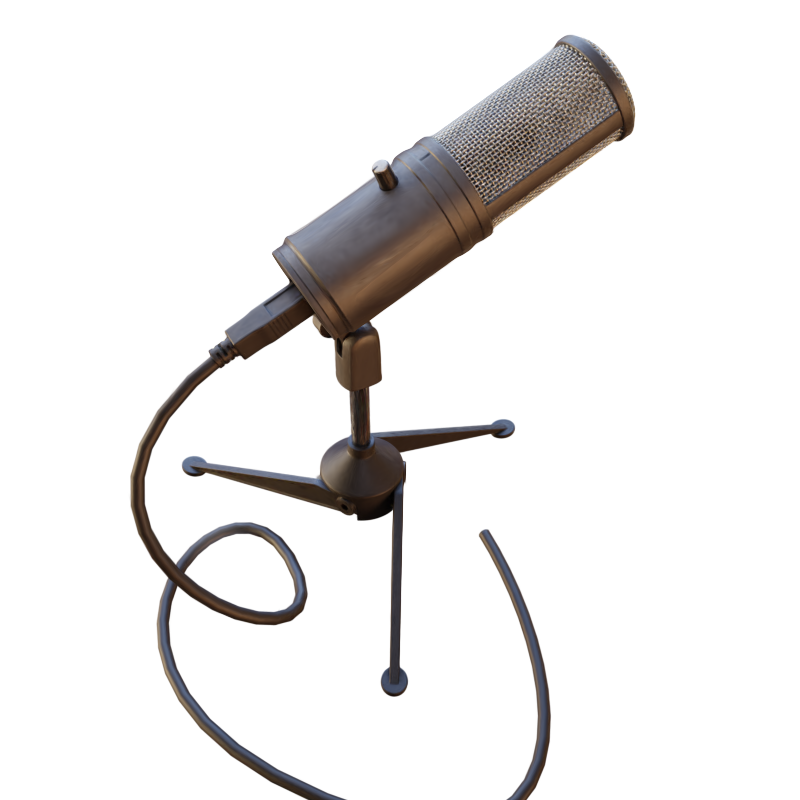} & \includegraphics[width=0.23\textwidth]{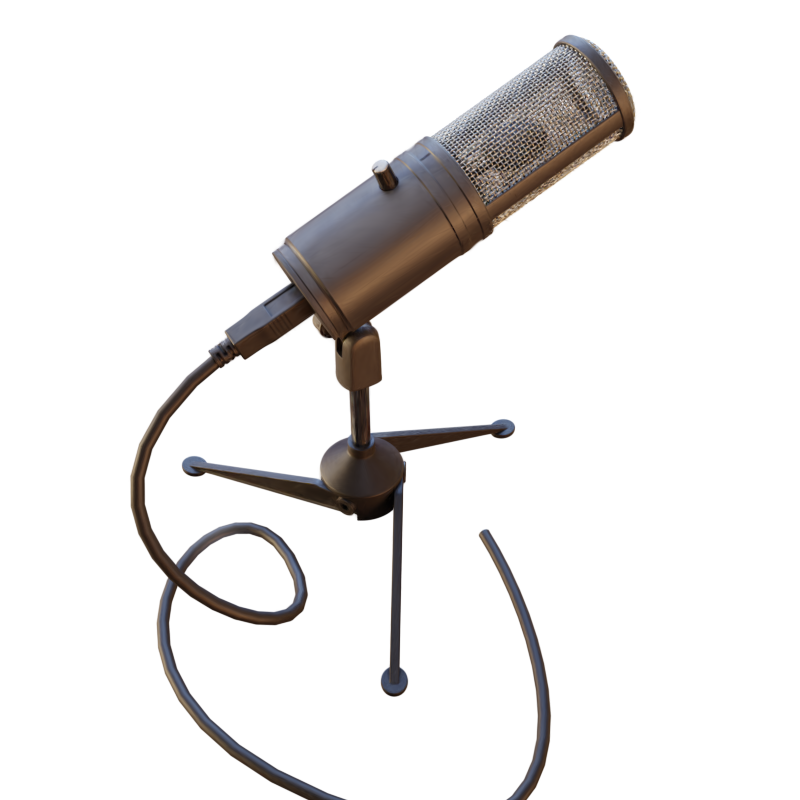} & \includegraphics[width=0.23\textwidth]{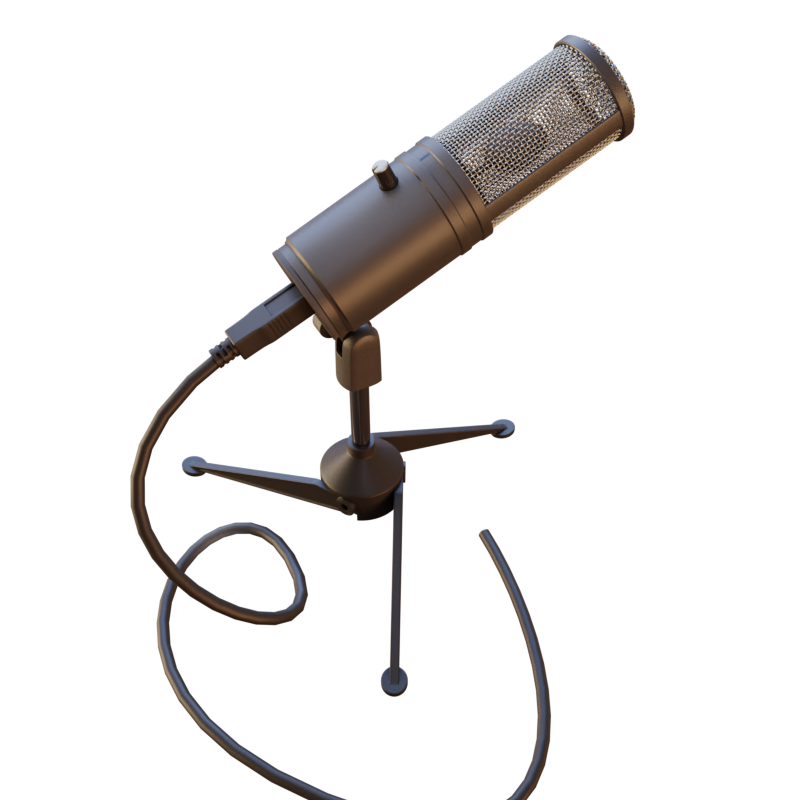}\\
        
        \rotatebox{90}{\hspace{2pt} Geometry} & \includegraphics[width=0.24\textwidth]{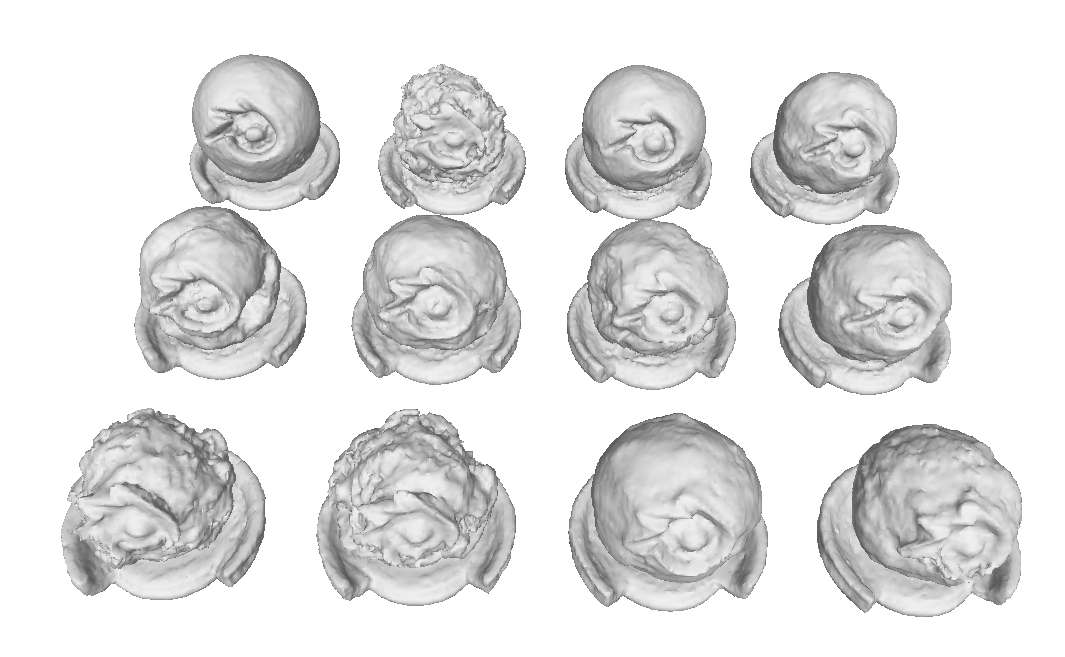} & \includegraphics[width=0.23\textwidth]{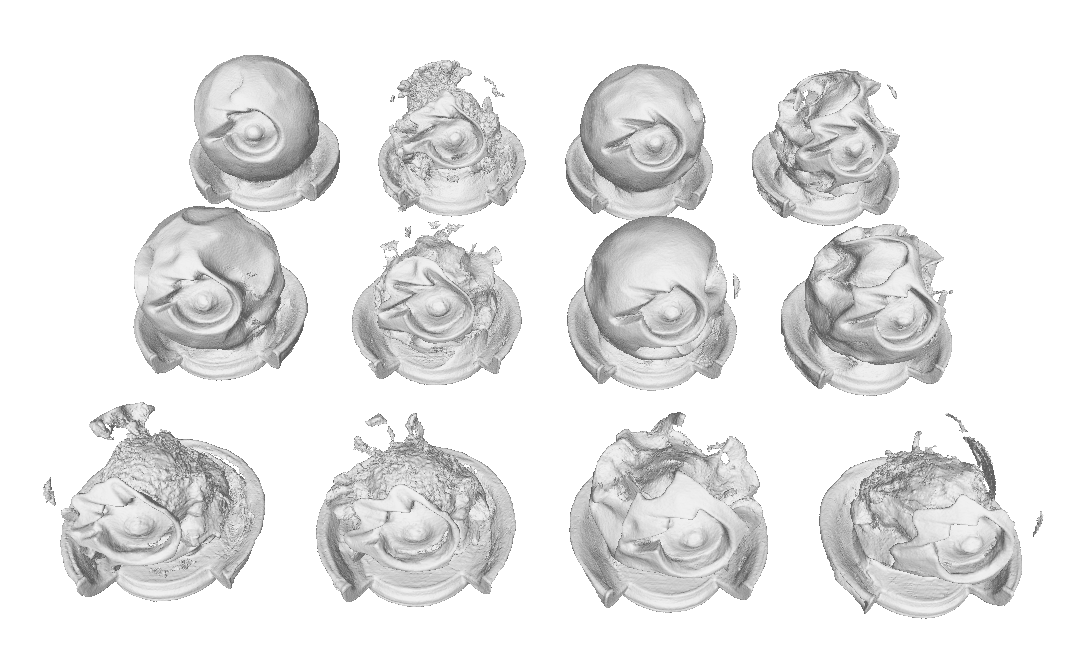} & \includegraphics[width=0.23\textwidth]{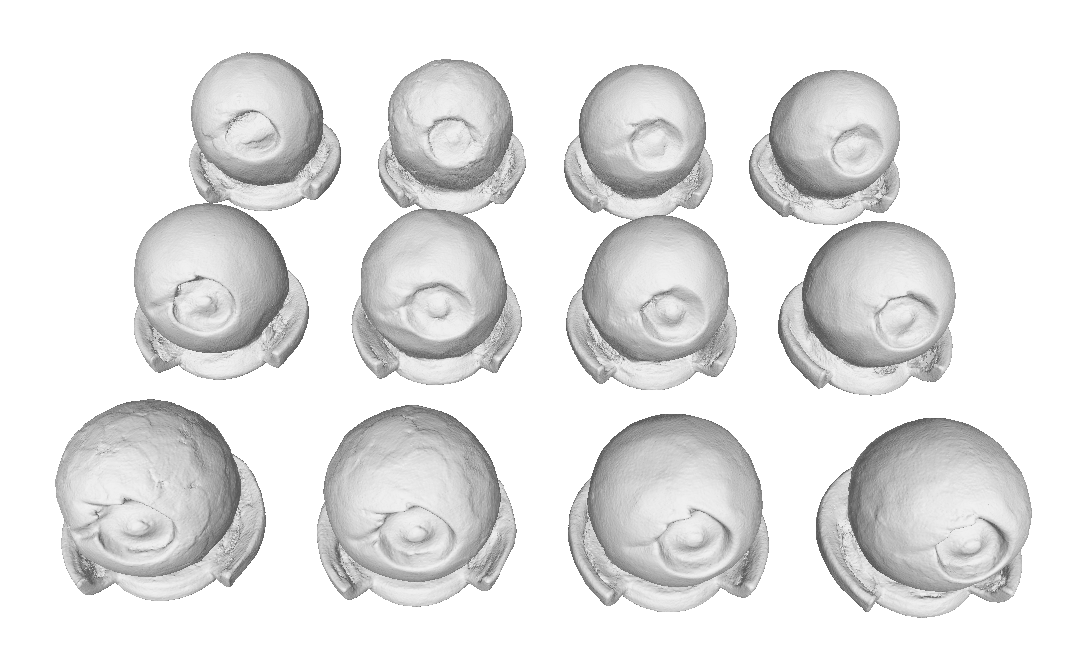} & \includegraphics[width=0.23\textwidth]{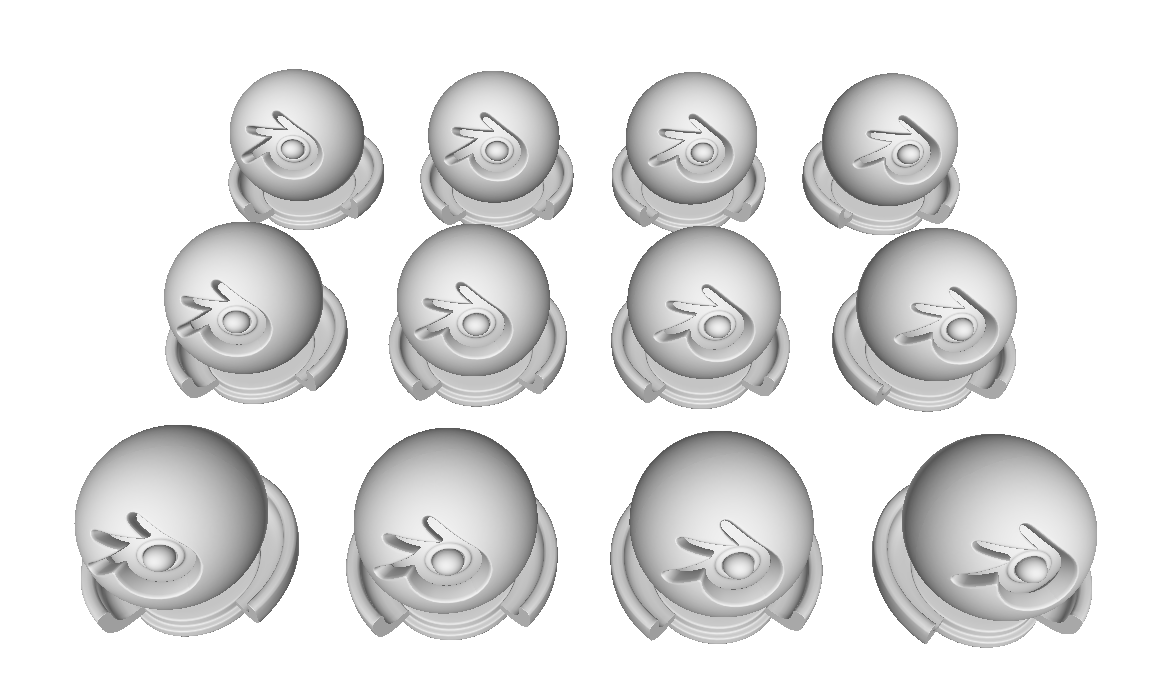}\\
        \rotatebox{90}{\hspace{10pt} Rendering} & \includegraphics[width=0.23\textwidth]{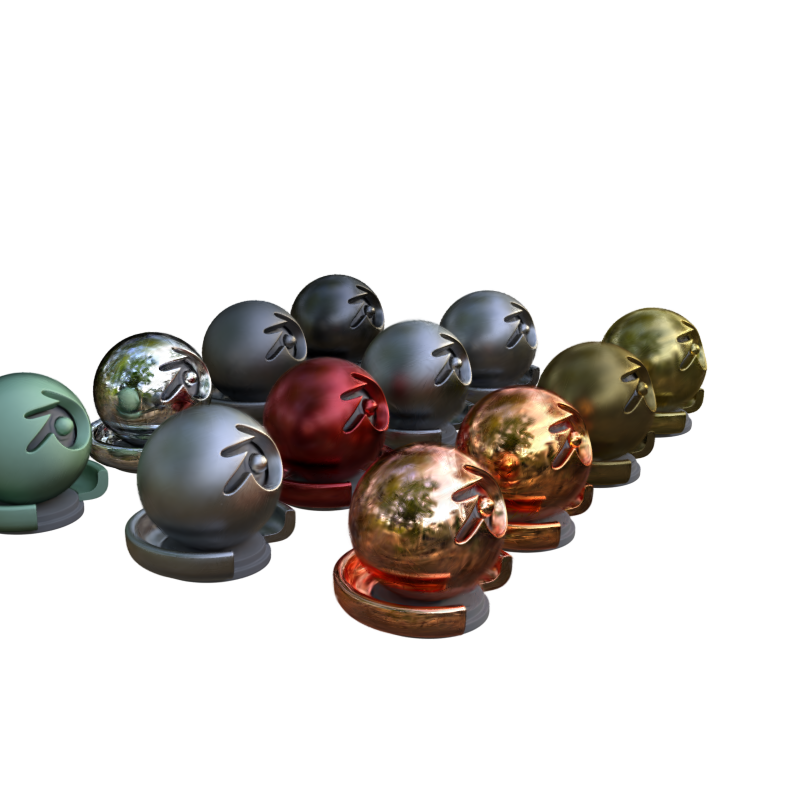} & \includegraphics[width=0.23\textwidth]{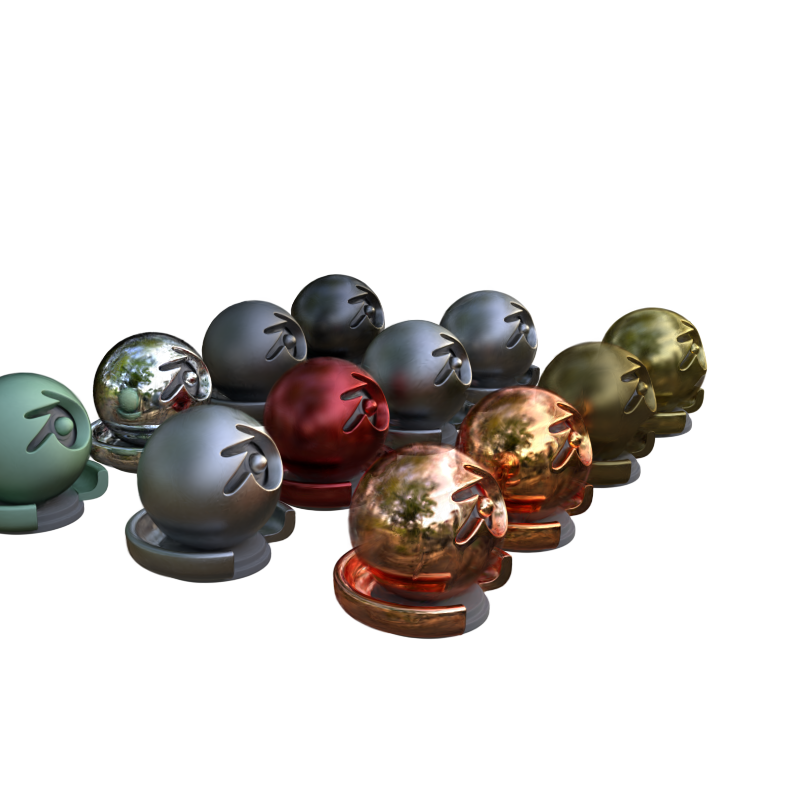} & \includegraphics[width=0.23\textwidth]{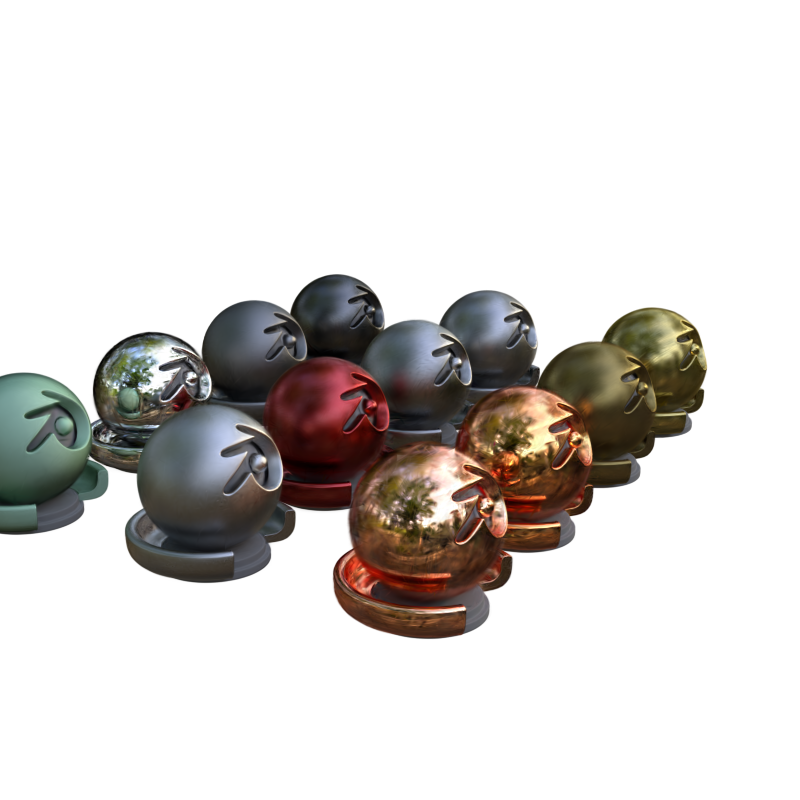} & \includegraphics[width=0.23\textwidth]{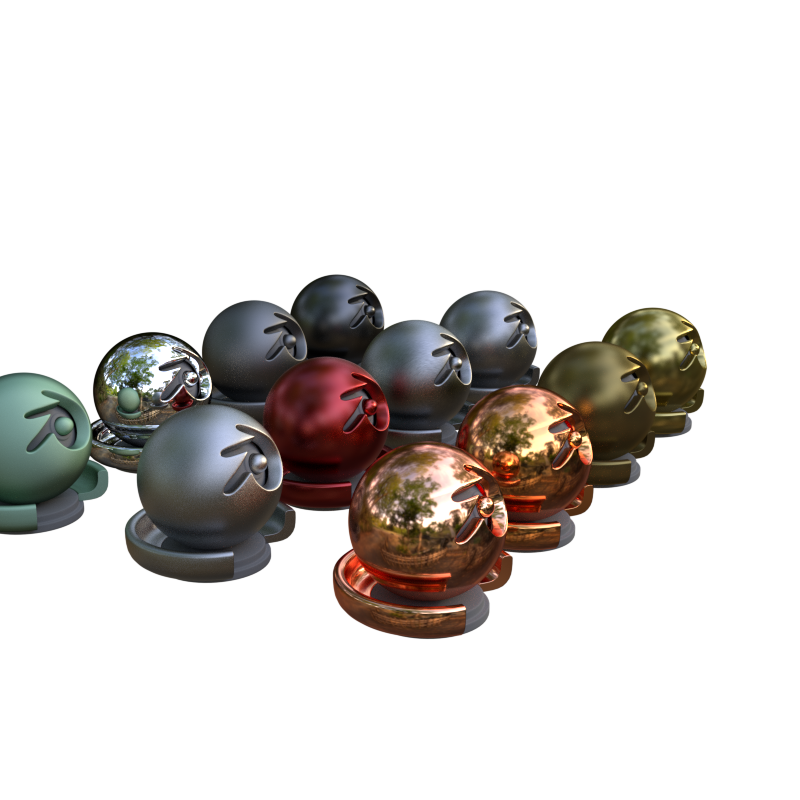} \\

    \end{tabular}
    \caption{Qualitative comparison on the NeRF Synthetic dataset. Our method reconstructs accurate geometry even in presence of highly reflective material which is not explicitly modeled in the pipeline. }
    \label{exp:qualitative:nerf}
\end{figure}

\begin{table}[tbh]
    \centering
    \begin{tabular}{l ccc c ccc c ccc}
    \toprule
        & \multicolumn{3}{c}{\text{Indoor}} & & \multicolumn{3}{c}{\text{Outdoor}} & & \multicolumn{3}{c}{\text{All Scenes}}\\
        \cmidrule{2-4}
        \cmidrule{6-8}
        \cmidrule{10-12}
        & PSNR$\uparrow$ & SSIM$\uparrow$ & LPIPS$\downarrow$ & & PSNR$\uparrow$ & SSIM$\uparrow$ & LPIPS$\downarrow$ & & PSNR$\uparrow$ & SSIM$\uparrow$ & LPIPS$\downarrow$\\
        \midrule
        3DGS~\cite{kerbl20233d} & 30.41 & 0.920 & 0.193 & &  \cellcolor{yellow!20}24.64 & 0.731 & 0.234 & & 27.21 & 0.815 & 0.216 \\
        
        2DGS~\cite{Huang2DGS2024} & 30.41 & 0.916 & 0.194 & & 24.34 & 0.717 & 0.246 & & 27.03 & 0.805 & 0.223 \\
        
        GOF~\cite{yu2024gaussian} & \cellcolor{orange!20}30.79 & 0.924 & 0.184 & & \cellcolor{red!20}24.82 &\cellcolor{orange!20}0.750 & \cellcolor{red!20}0.202 & & \cellcolor{orange!20}27.47 & \cellcolor{orange!20}0.827 &  \cellcolor{yellow!20}0.194 \\
        
        PGSR~\cite{chen2024pgsr} & 30.36 & \cellcolor{red!20}0.934 & \cellcolor{red!20}0.147 & & \cellcolor{orange!20}24.76 & \cellcolor{red!20}0.752 & \cellcolor{orange!20}0.203 & & 27.25 & \cellcolor{red!20}0.833 & \cellcolor{red!20}0.178 \\
        
        CarGS~\cite{shen2025evolving}$^{\star}$ & \cellcolor{yellow!20}30.42 &  \cellcolor{yellow!20}0.928 &  \cellcolor{yellow!20}0.171 & & 24.51 &  \cellcolor{yellow!20}0.741 & 0.246 & & \cellcolor{yellow!20}27.33 &  \cellcolor{yellow!20}0.824 & 0.213 \\
        
        Ours & \cellcolor{red!20}31.21 & \cellcolor{orange!20}0.933 &  \cellcolor{orange!20}0.148 & &  \cellcolor{yellow!20}24.64 & 0.731 &  \cellcolor{yellow!20}0.207 & &\cellcolor{red!20}27.56 & 0.821 & \cellcolor{orange!20}0.181 \\
    \bottomrule
    \end{tabular}
    \caption{Rendering results on MipNeRF-360. Our method is on-par with PGSR and GOF. ${\star}$ results are copied from the original paper since no implementation is available. (best result in red, second best in orange, third best in yellow)}
    \label{exp:quantitative:Mipnerf}
\end{table}
\paragraph{Transparency.} Results for novel-view rendering and reconstruction are shown in \cref{exp:quantitative:translab}. 
We compare our method with 2DGS~\cite{Huang2DGS2024}, PGSR~\cite{chen2024pgsr} and TSGS~\cite{li2025tsgs}, the SOTA method for transparent object reconstruction from Gaussian Splatting. 
Our method achieves best reconstruction with regard to Chamfer distance and F1-score with a distance threshold equal to 0.005, and the best quality of novel view rendering. 
In addition, our method is 3 times faster than the direct competitor TSGS. Qualitative result can be found in \cref{exp:qualitative:translab} where our method shows significant improvement in rendering quality. 
\paragraph{Reflectance.} We compare our method with 2DGS\footnote{\small{Note that the default script provided by 2DGS for the NeRF Synthetic dataset does not have geometric regularization.}} and PGSR on the NeRF-Synthetic dataset in novel-view rendering and reconstruction in \cref{exp:quantitative:translab}. 
Our method achieves results on-par with the best performance in both rendering and reconstruction but the qualitative examples in \cref{exp:qualitative:nerf} show significantly better reconstruction quality under strong light effects.
\paragraph{Ordinary Materials.} Additionally, we evaluate our method on DTU and MipNeRF-360, which do not contain large quantities of complex materials. 
The results can be seen in \cref{exp:quantitative:DTU} and \cref{exp:quantitative:Mipnerf}. 
Qualitative examples are shown in the supplementary.
For DTU, which is only about reconstruction, we compare with 2DGS, GOF, and PGSR, which are methods focused on geometry. 
Our method has the second best performance on DTU, slightly behind PGSR. 
For MipNeRF-360, which is for rendering only, we compare with 3DGS, 2DGS, GOF, PGSR, and CarGS, and we achieve the best performance, especially for indoor scenes. 
The best performances on these datasets are quite similar and we can see that geometry opacity is not a large advantage here, but it also does not degrade the results which makes it widely applicable even without prior knowledge what kind of scene is handled. 

\begin{figure}[tb]
    \centering
    \begin{tabular}{ccccccc}
        Full & -Opacity$_{geo}$ & -VGGT & -DKT & -Offset & -MVS mask & GT\\
        \includegraphics[width=0.13\textwidth]{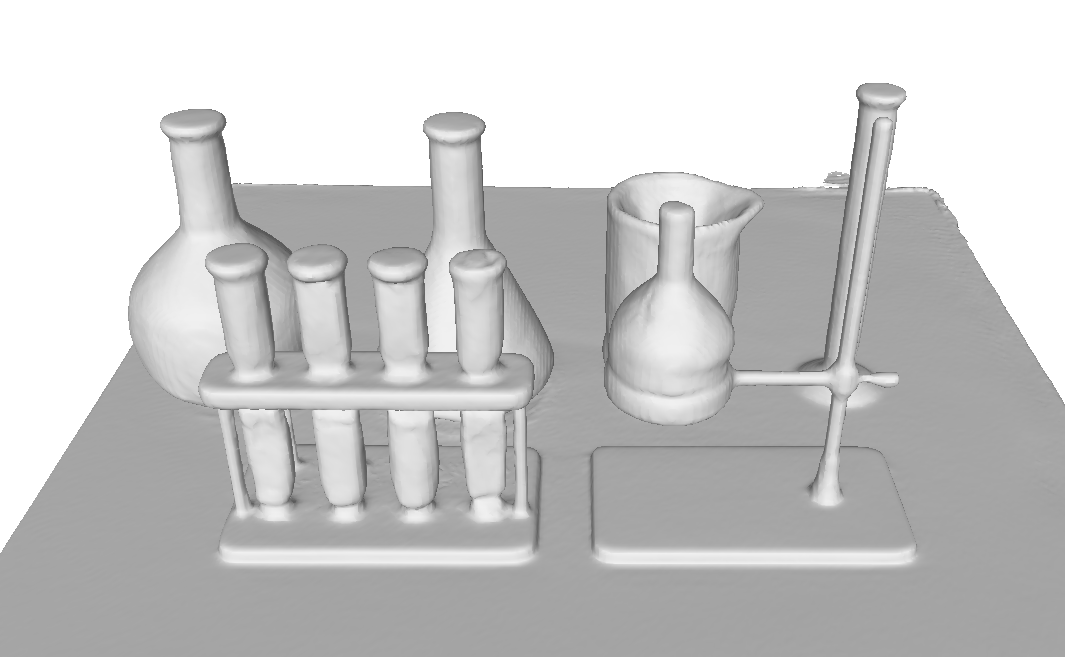} & \includegraphics[width=0.13\textwidth]{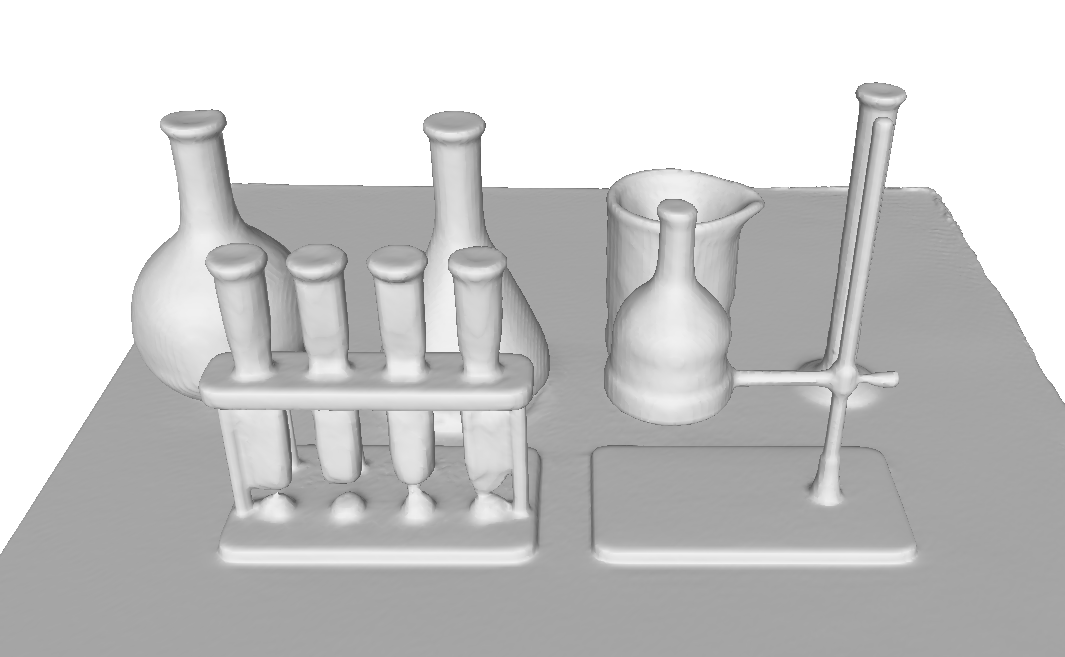} & \includegraphics[width=0.13\textwidth]{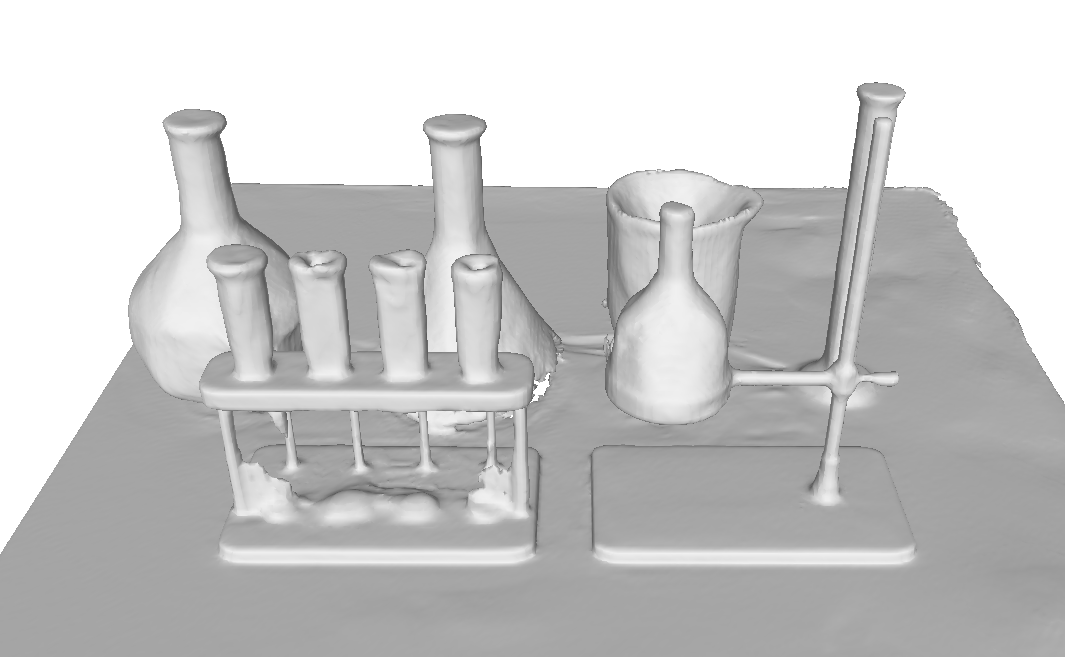} & \includegraphics[width=0.13\textwidth]{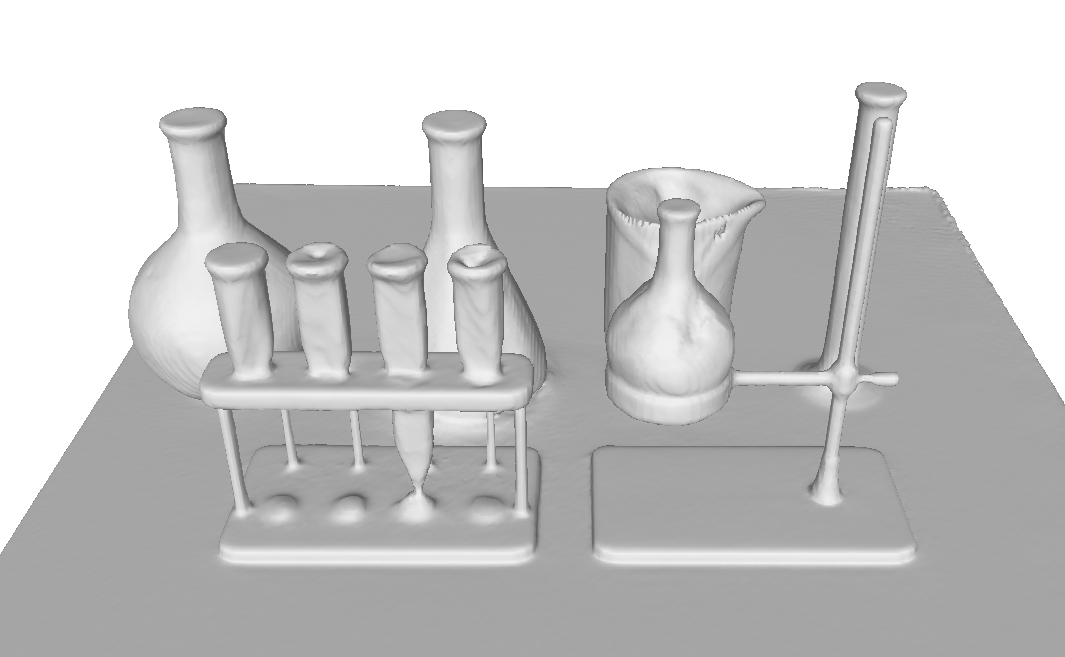} & \includegraphics[width=0.13\textwidth]{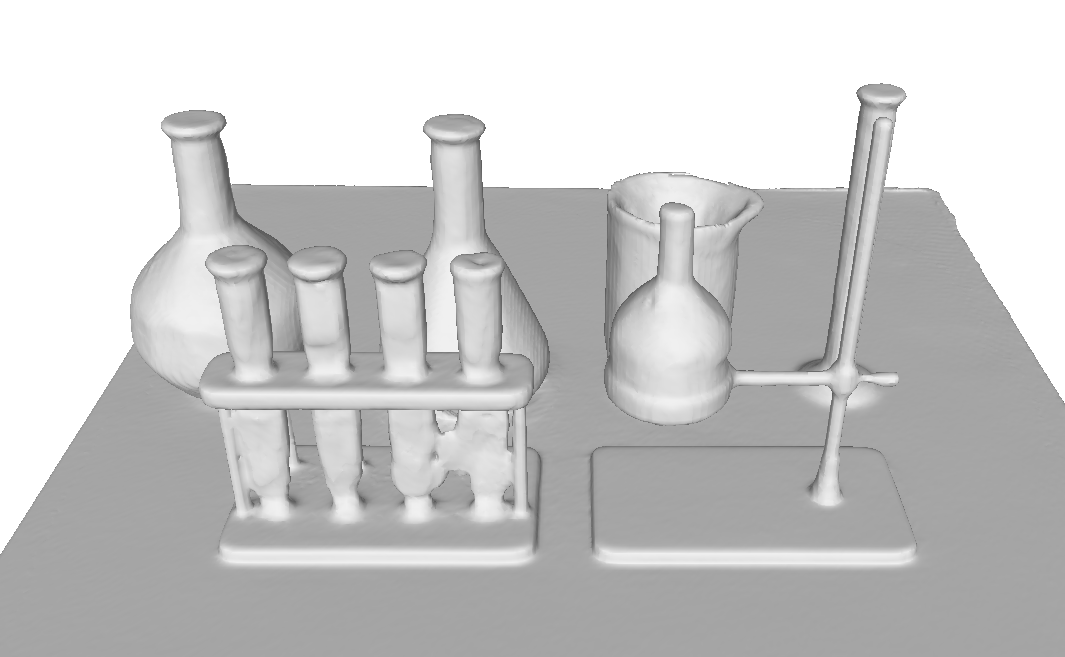} & \includegraphics[width=0.13\textwidth]{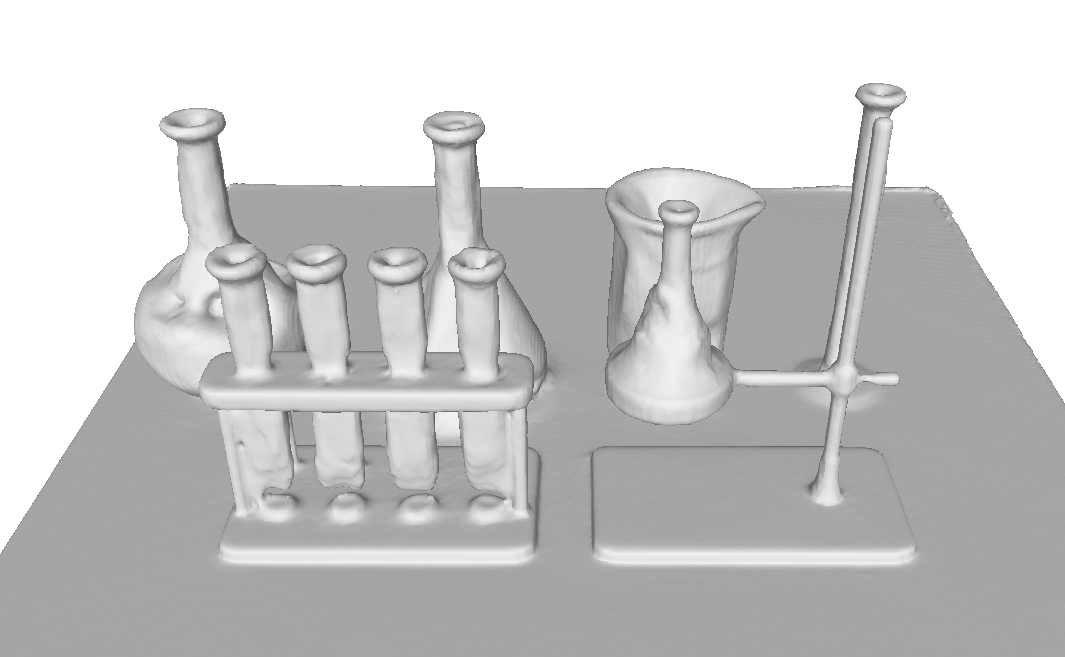} & \includegraphics[width=0.13\textwidth]{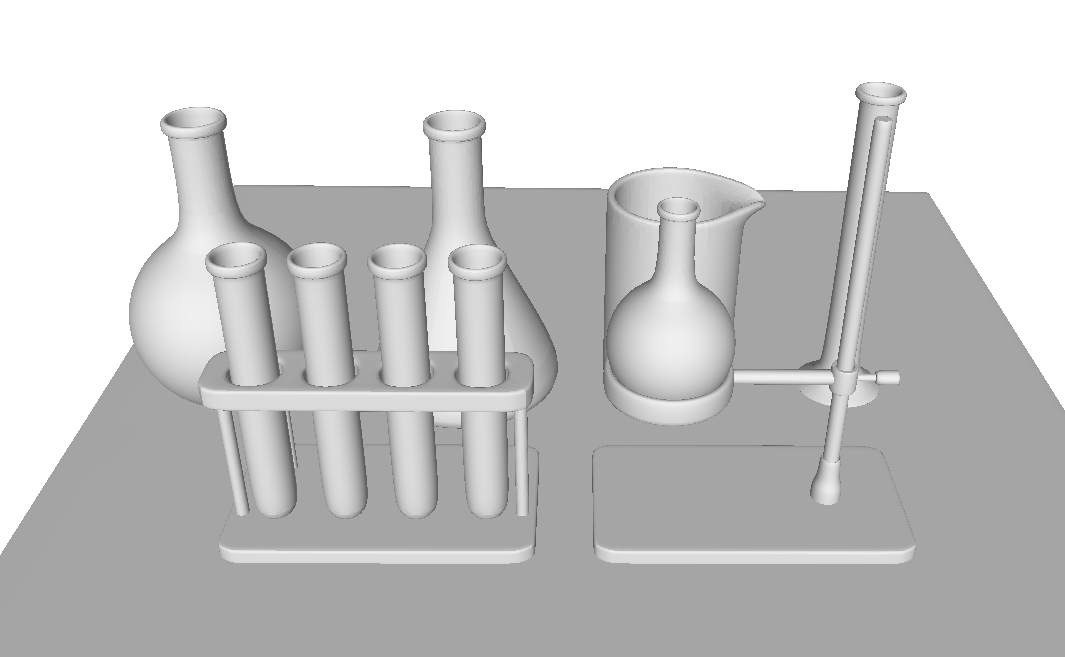} \\   
    \end{tabular}
    \caption{Qualitative examples for the ablation study. Even though removing DKT leads to the lowest Chamfer distance in \cref{exp:ablation:quantitative}, it is clearly visible that the reconstruction of the test tubes is insufficient. }
    \label{exp:ablation:qualitative}
\end{figure}

\begin{table}[tbh]
    \centering
    \resizebox{\textwidth}{!}{
    \begin{tabular}{l|c c c c c c c c c c c c c c c|c}
    \toprule
          \textbf{DTU} & 24 & 37 & 40 & 55 & 63 & 65 & 69 & 83 & 97 & 105 & 106 & 110 & 114 & 118 & 122 & Mean\\ 
          \midrule
         2DGS~\cite{Huang2DGS2024} & 0.48 & 0.91 & 0.39 & 0.39 & 1.01 & 0.83 & 0.81 & 1.36 & 1.27 & 0.76 & 0.70 & 1.40 & 0.40 & 0.76 & 0.52 & 0.80 \\ 
         GOF~\cite{yu2024gaussian} & 0.50 & 0.82 & 0.37 & 0.37 & 1.12 & 0.74 & 0.73 & 1.18 & 1.29 & 0.68 & 0.77 & 0.90 & 0.42 & 0.66 & 0.49 & 0.74 \\
         PGSR~\cite{chen2024pgsr} & 0.36 & \cellcolor{red!20}0.57 & 0.38 & \cellcolor{red!20}0.33 & \cellcolor{red!20}0.78 & \cellcolor{red!20}0.58 & \cellcolor{red!20}0.50 & \cellcolor{red!20}1.08 & \cellcolor{red!20}0.63 & \cellcolor{red!20}0.59 & 0.46 &  \cellcolor{red!20}0.54 & \cellcolor{red!20}0.30 & \cellcolor{red!20}0.38 &\cellcolor{red!20}0.34 & \cellcolor{red!20}0.52 \\
         % + SN & 0.40 & 0.52 & 0.35 & 0.35 & 0.79 & 0.57 & 0.53 & \textbf{1.06} & 0.70 & 0.60 & 0.42 & 0.54 & 0.31 & 0.40 & 0.35 & 0.53 \\
         Ours & \cellcolor{red!20}0.34 & \cellcolor{orange!20}0.69 & \cellcolor{red!20}0.33 & \cellcolor{orange!20}0.34 & \cellcolor{orange!20}0.88 & \cellcolor{orange!20}0.60 & \cellcolor{red!20}0.50 & \cellcolor{orange!20}1.14 & \cellcolor{orange!20}0.76 & \cellcolor{orange!20}0.61 & \cellcolor{red!20}0.43 & \cellcolor{orange!20}0.56 & \cellcolor{red!20}0.30 & \cellcolor{orange!20}0.45 & \cellcolor{orange!20}0.38 & \cellcolor{orange!20}0.55 \\
    \bottomrule
    \end{tabular}
    }
    \caption{Chamfer distance on DTU. Our method performs similarly to PGSR. (best result in red, second best in orange)}
    \label{exp:quantitative:DTU}
\end{table}

\begingroup
\subsection{Ablation Study}\label{sub:ablation}
\setlength\intextsep{0pt}
\begin{wraptable}[11]{r}{0.6\textwidth}
\centering
\resizebox{\linewidth}{!}{
    \centering
    \begin{tabular}{l c c c} 
    \toprule
       TransLab  &  PSNR$\uparrow$ & CD$\downarrow$ & F1$\uparrow$ \\ 
         \midrule
     \text{Full model}& \cellcolor{yellow!20}$39.95\pm3.25\ $ & \cellcolor{orange!20}$1.665\pm0.266\ $& \cellcolor{red!20}$0.960\pm0.019$ \\
     \text{- Opacity$_{geo}$} & $37.81\pm3.08$ & \cellcolor{yellow!20}$1.679\pm0.293$ & \cellcolor{yellow!20}$0.957\pm0.019$  \\
     \text{- VGGT} & $39.88\pm3.06$ & $2.245\pm 0.311$ & $0.920\pm 0.036$ \\
     \text{- DKT} & \cellcolor{red!20}$40.08\pm3.47$ & \cellcolor{red!20}$1.611\pm0.427$ & \cellcolor{yellow!20}$0.957\pm0.027$ \\
     \text{- Offset learning} & $39.65\pm3.23$ & $1.671\pm0.312$ & \cellcolor{orange!20}$0.959\pm0.023$ \\
     % \text{-Segmentation Learning} & 39.33 & 1.714 & 0.959 \\
     \text{- MVS mask} & \cellcolor{orange!20}$40.03\pm 3.43$ & $1.833\pm0.412$ & $0.938\pm0.034$\\
     \bottomrule
    \end{tabular}
    }
    \caption{Quantitative results of the ablation study. (best result in red, second best in orange, third best in yellow.)}
    \label{exp:ablation:quantitative}
\end{wraptable}
We conduct an ablation study on the TransLab dataset by removing every part of our pipeline. 
In addition to the average performance, we also evaluate its robustness in different scenes by comparing their standard deviation, as shown in \cref{exp:ablation:quantitative}. Qualitative results are given in \cref{exp:ablation:qualitative}.

\noindent\textbf{Geometry Opacity. } As pointed out in \cref{subsec:rep}, geometry opacity can enhance the representation power of Gaussian Splatting, so the rendering quality decreases when the geometry opacity is removed. In \cref{exp:ablation:qualitative} and \cref{exp:ablation:quantitative}, we show that the reconstruction is not as affected by the removal.
However, both cannot be optimized well with a single opacity.

\noindent\textbf{Detection-guided supervision. } When geometric supervision from open-domain perception models (\ie VGGT) is removed, the reconstruction quality drops significantly. Although geometric supervision from a fine-tuned model (\ie DKT) does not change the quality in average, it is more vulnerable to the scene because the geometry of the transparent part completely relies on the rendering loss that changes the position. The geometry opacity for these splats will remain unchanged. In \cref{exp:ablation:qualitative}, the tube is missing and there is no way to retrieve it.
\endgroup

\noindent\textbf{Transparency-aware Multi-view Stereo Loss. } \cref{exp:ablation:qualitative} and \cref{exp:ablation:quantitative} show that the mask in the photometric multi-view stereo loss is essential to prevent the shape of transparent objects to shrink, caused by false photometric clues.

\section{Conclusion} \label{sec:conclusion}

We introduced \emph{geometry opacity}, a novel way to disentangle rendering and geometric properties in Gaussian splatting by adding a single parameter to each splat. 
Our experiments with complete ground-truth information during training show clearly that the default 3DGS parametrization is not able to represent color and geometry information perfectly at the same time, while our light-weight change allows both with much less overhead then keeping a completely separate geometry representation, like a signed distance function.
This is especially advantageous for objects with transparency.
To incorporate geometry opacity in real-world settings, we proposed additional regularizers that can be used with geometry-aware 3DGS pipelines to make them more robust to errors and misalignment that often happen with vision foundation models when transparency is present in the scene. 
Our proposed method clearly outperforms default 3DGS setups on data with transparency and does not degrade performance in more default cases, which we showed on the NeRF Synthetic, DTU, Translab and Mip-NeRF datasets. 
At the same time, it is light-weight and flexible enough to be included in any geometry-aware 3DGS pipeline and, thus, applicable in many different settings.

%\section*{Acknowledgements}
%Please insert your acknowledgments here.

\bibliographystyle{splncs04}
\bibliography{main}
\end{document}